\documentclass[12pt]{article}
\catcode`\@=11
\@addtoreset{equation}{section}

\global\arraycolsep=2pt
\usepackage{mathrsfs,amsbsy,amssymb,latexsym,amsfonts,amsmath,cite}
\usepackage{graphicx,color}
\usepackage{physics}

\usepackage{mathtools}
\usepackage{ulem}

\newcommand{\pa}{\partial}

\newcommand{\cc}[1]{\overline{#1}}

\newcommand{\no}{\nonumber}

\newcommand{\cE}{\mathcal E}\newcommand{\cF}{\mathcal F}

\newcommand{\cL}{\mathcal L}
\newcommand{\cM}{\mathcal M}
\newcommand{\cO}{\mathcal O}

\newcommand{\sfL}{\mathsf L}

\newcommand{\sfR}{\mathsf R}

\newcommand{\zR}{z_{\mathsf{R}}}

\newcommand{\laR}{\lambda_{\mathsf{R}}}
\newcommand{\wR}{w_{\mathsf{R}}}
\newcommand{\zL}{z_{\mathsf{L}}}
\newcommand{\zLp}{z_{\mathsf{L}_{+}}}
\newcommand{\zLm}{z_{\mathsf{L}_{-}}}
\newcommand{\zLpm}{z_{\mathsf{L}_{\pm}}}

\newcommand{\Se}{\Sigma^{(\mathrm{e})}}\newcommand{\So}{\Sigma^{(\mathrm{o})}}
\newcommand{\Ze}{\mathcal{Z}^{(\mathrm{e})}}\newcommand{\Zo}{\mathcal{Z}^{(\mathrm{o})}}
\newcommand{\Ee}{\mathcal{E}^{(\mathrm{e})}}\newcommand{\Eo}{\mathcal{E}^{(\mathrm{o})}}
\newcommand{\Fe}{\mathcal{F}^{(\mathrm{e})}}\newcommand{\Fo}{\mathcal{F}^{(\mathrm{o})}}

\newcommand{\llangle}{ \langle\!\langle}
\newcommand{\rrangle}{ \rangle\!\rangle}

\allowdisplaybreaks

\begin{document}

\begin{flushright}
KUNS-2802
\end{flushright}
\vspace*{0.5cm}

\begin{center}
{\Large \bf Comments on $\eta$-deformed principal chiral model \\[5pt]
from 4D Chern-Simons theory}
\vspace*{1.5cm} \\
{\large  Osamu Fukushima$^{\sharp}$\footnote{E-mail:~osamu.f@gauge.scphys.kyoto-u.ac.jp},
Jun-ichi Sakamoto$^{\dagger,\ddagger}$\footnote{E-mail:~sakamoto@ntu.edu.tw},
and Kentaroh Yoshida$^{\sharp}$\footnote{E-mail:~kyoshida@gauge.scphys.kyoto-u.ac.jp}} 
\end{center}

\vspace*{0.4cm}

\begin{center}
$^{\sharp}${\it Department of Physics, Kyoto University, Kyoto 606-8502, Japan}
\end{center}
\begin{center}
$^{\dagger}${\it Department of Physics and Center for Theoretical Sciences, National Taiwan University, Taipei 10617, Taiwan}
\end{center}
\begin{center}
$^\ddagger${\it Osaka City University Advanced Mathematical Institute (OCAMI), 3-3-138, Sugimoto, Sumiyoshi-ku, Osaka, 558-8585, Japan}
\end{center}

\vspace{1cm}

\begin{abstract}
We study $\eta$-deformations of principal chiral model (PCM) 
from the viewpoint of a 4D Chern-Simons (CS) theory.  
The $\eta$-deformed PCM has originally been derived from the 4D CS theory 
by Delduc, Lacroix, Magro and Vicedo [arXiv:1909.13824]. The derivation is based 
on a twist function in the rational description. On the other hand, we start with 
a twist function in the trigonometric description and discuss possible boundary conditions. 
We show that a certain boundary condition reproduces the usual $\eta$-deformed PCM 
and another one leads to a new kind of Yang-Baxter deformation.  
\end{abstract}

\setcounter{footnote}{0}
\setcounter{page}{0}
\thispagestyle{empty}

\newpage

\tableofcontents

\section{Introduction}

Classically integrable field theories provide a good arena for examining non-linear dynamics. 
It is a significant direction to consider a systematic way to construct the integrable 
field theories \cite{CWY1,CWY2,CY}. Recently, Costello and Yamazaki \cite{CY} made 
an interesting proposal along this line. According to it, starting from a certain 
4D Chern-Simons (CS) theory, one can construct classically integrable field theories 
systematically by taking a meromorphic 1-form and adopting an appropriate 
boundary condition. In other words, the choice of the meromorphic 1-form and 
boundary condition determines the associated integrable field theory. 

\medskip 

On the other hand, some techniques to perform integrable deformation are also useful 
for generating new integrable field theories. For example, there has been much progress 
for systematic ways to discuss integrable deformations of 2D non-linear sigma model, 
such as the Yang-Baxter deformation \cite{Klimcik1,Klimcik2} 
and the $\lambda$-deformation \cite{lambda1,lambda2}. 
The Yang-Baxter (YB) deformation was originally invented for 2D principal chiral model (PCM) 
with the modified classical Yang-Baxter equation (mCYBE) \cite{Klimcik1,Klimcik2}
and then generalized to the symmetric coset case \cite{DMV-symm,DMV2-IIB} 
and to the homogeneous classical Yang-Baxter equation (hCYBE) \cite{KMY1, Matsumoto:2015jja}. 
In particular, the YB deformation based on the mCYBE is often called the $\eta$-deformation.  

\medskip

In the very recent, Delduc, Lacrox, Magro and Vicedo succeeded in discussing the YB deformation 
\cite{DLMV} along with the Costello-Yamazaki proposal \cite{CY}. A profound discovery made in \cite{Gaudin} 
is that the meromorphic function is nothing but a twist function characterizing 
the classical integrable structure. That is, by starting with the associated twist function, 
the meromorphic 1-form in \cite{CY} is automatically determined. 
Then one can figure out the well-known integrable deformations 
as the associated boundary conditions. 

\medskip 

In this paper, we are concerned with a realization of the $\eta$-deformation of 2D PCM 
in \cite{DLMV}. Assume that the dynamical variable $g$ of PCM takes a value 
in a Lie group $G$. Then the PCM has the left and right symmetries, $G_L$ and $G_R$\,, respectively. 
Under the $\eta$-deformation, one of them is broken to $U(1)^r$\,, where $r$ is the rank of $G$\,. 
In our later discussion, we will suppose that $G_R$ is broken 
while $G_L$ remains unbroken. The resultant $U(1)^r$ symmetry can be regarded as 
the level zero part of an affine extension of $q$-deformed $G_R$\,, $\widehat{U}_q(\mathfrak{g}^R)$ 
\cite{KMY-cQA,DMV-symm}, 
while the unbroken $G_L$ is enhanced to the Yangian algebra $Y(\mathfrak{g}^L)$ \cite{KY-Yangian,KMY-cQA}. 

\medskip

It is remarkable that the left-right duality is still realized in a non-trivial way 
even after performing the $\eta$-deformation \cite{KMY-monod} in the $\mathfrak{su(2)}$ case. 
According to this duality, there are two manners to 
describe the dynamics of the $\eta$-deformed PCM, 1) the trigonometric description (based 
on $\widehat{U}_q(\mathfrak{g}^R)$) and 2) the rational description (based on $Y(\mathfrak{g}^L)$), 
and the two descriptions are equivalent under a certain relation of spectral parameters 
\cite{KMY-monod}. According to the two ways, one may consider two kinds of twist functions.  
In the work \cite{DLMV}, a twist function in the rational description is utilized. 

\medskip 

Our purpose here is to revisit the $\eta$-deformed PCM by starting 
with the trigonometric description. Then the spectral parameter takes a value 
on a cylinder rather than a sphere. This cylinder is equivalent to a couple of spheres 
and the two descriptions are equivalent by taking account of the gauge-transformed monodromy 
in the rational description.
On the other hand, by starting from the trigonometric description, 
we can discuss the whole space of spectral parameter by construction all at once, 
and hence we could not only reproduce 
the usual results on the $\eta$-deformed PCM, but also discover a new type of 
YB deformation as a byproduct. This is the main result of our paper.  

\medskip 

This paper is organized as follows. 
Section 2 gives a short review of the work \cite{DLMV} 
by focusing upon the $\eta$-deformed PCM. 
In Section 3, we study the $\eta$-deformed PCM  
by employing the trigonometric description. 
In particular, the range of spectral parameter becomes twice 
in comparison to the analysis in Section 2.  Two boundary conditions lead to 
the usual results and a new type of YB deformation, respectively. 
In Section 4, the left-right duality is discussed in the $\eta$-deformed PCM.
Section 5 is devoted to conclusion and discussion. Appendix A explains the details on how to find 
appropriate boundary conditions. Appendix B explains how to realize 
a possible $\lambda$-map. 
Appendix C gives a direct proof of the integrability of the new-type of YB deformed sigma model. 
In Appendix D, we discuss a specialty of the $SU(2)$ case. 
In particular, the two boundary conditions are related by a singular gauge transformation.  
Appendix E discusses a scaling limit of the $\eta$-deformed 
$SL(2,\mathbb{R})$ PCM at the level of twist function.

\section{$\eta$-deformed PCM from 4D CS theory} 

This section provides a short review on a procedure to derive 2D 
integrable sigma models from a 4D CS theory \cite{CY,DLMV} 
and describes how to derive the $\eta$-deformed PCM by the {\it rational} 
description\footnote{One may begin with the {\it trigonometric} description 
and this will be the subject of Section \ref{trigonometric description}.}.  

\subsection{4D CS action}

Let $G^{\mathbb{C}}$ be a complexified semisimple Lie group with Lie algebra 
$\mathfrak{g}^{\mathbb{C}}$ equipped with a non-degenerate symmetric bilinear form 
$\langle\cdot,\cdot\rangle:\mathfrak{g}^{\mathbb{C}}\times\mathfrak{g}^{\mathbb{C}}
\rightarrow\mathbb{C}$\,. The bilinear form $\langle\cdot,\cdot\rangle$ is also adjoint-invariant:
\begin{align}
\langle B,[C,D]\rangle=-\langle [C,B],D\rangle\,.
\end{align}
In the following, we will consider a $\mathfrak{g}^{\mathbb{C}}$-valued gauge field $A$ defined on 
$\cM \times \mathbb{C}P^1$\,. Here $\cM$ is a 2D Minkowski space with the coordinates 
$x^{i} = (x^0,x^1)=(\tau,\sigma)$ and the metric is given by $\eta_{ij} = \mbox{diag}(-1,+1)$\,. 
The global holomorphic coordinate of $\mathbb{C}P^1 :=\mathbb{C}\cup\{\infty\}$ 
is denoted by $z$\,. This $\mathbb{C}P^1$ geometry characterizes the rational class 
of integrable system. 

\medskip

By following \cite{CY}, we shall begin with a 4D CS action,  
\begin{align}
S[A]=-\frac{i}{4\pi}\int_{\cM\times\mathbb{C}P^1} \omega\wedge CS(A)\,. 
\label{4dcs}
\end{align}
Here $\omega$ is a meromorphic 1-form defined as 
\begin{align}
\omega:=\varphi(z)dz\,, \label{omega}
\end{align}
where $\varphi$ is a meromorphic function defined on $\mathbb{C}P^1$\,. Remarkably, 
this function has been identified with a twist function characterizing the Poisson structure 
of the underlying integrable field theory \cite{DLMV}.  

\medskip

In the following discussion, the pole and zero structure of $\varphi$ will play a significant role. 
The set of poles and zeros of $\varphi$ is denoted as $\mathfrak{p}$ and $\mathfrak{z}$\,, 
respectively. 

\medskip

As usual, the CS 3-form is defined as 
\begin{align}
CS(A):= \left\langle A,dA+\frac{2}{3}A\wedge A\right\rangle\,,
\end{align}
where $A$ is a $\mathfrak{g}^\mathbb{C}$-valued 1-form 
\begin{align}
A=A_\sigma\, d\sigma+A_\tau\, d\tau+A_{\bar{z}}\, d\bar{z}\,.
\end{align}
Note here that the $z$-component can always be ignored 
because the action (\ref{4dcs}) has an extra gauge symmetry
\begin{align}
A\mapsto A+\chi\, dz\,,\label{extra gauge}
\end{align}
because $\omega$ is a (1,0)-form, hence the gauge condition $A_z=0$ can be realized. 

\medskip 

With respect to the gauge field $A$\,, a variation of the action (\ref{4dcs}) is expressed as 
\begin{align}
\delta S[A]=-\frac{i}{2\pi}\int_{\cM\times \mathbb{C}P^1}\omega\wedge\langle \delta A,F(A)\rangle-\frac{i}{4\pi}\int_{\cM\times \mathbb{C}P^1}d\omega\wedge\langle A,\delta A\rangle\,,\label{S variation}
\end{align}
where the field strength $F(A):=dA+A\wedge A$ and 
$A$ is assumed to vanish at the boundary of $\cM\times\mathbb{C}P^1$\,. 
Then the variation (\ref{S variation}) indicates that the action (\ref{4dcs}) has stationary points 
specified by the bulk equation of motion  
\begin{align}
\omega\wedge F(A)=&0\,,\label{bulk eom}
\end{align}
and the boundary equation of motion 
\begin{align}
d\omega \wedge \langle A,\delta A\rangle=&0\,.\label{boundary eom}
\end{align}
Note that the boundary equation of motion (\ref{boundary eom}) has the support only on 
$\cM\times\mathfrak{p}\subset \cM\times \mathbb{C}P^1$\,, because 
\[
d \omega = \partial_{\bar{z}}\varphi(z) d\bar{z} \wedge dz 
\]
and only the pole of $\varphi$ can contribute as a distribution. 
The boundary conditions satisfying (\ref{boundary eom}) are crucial to describe integrable deformations\cite{DLMV}.

\medskip

The bulk equation of motion (\ref{bulk eom}) can be expressed  in terms of the component fields:
\begin{align}
\partial_\sigma A_\tau -\partial_\tau A_\sigma+[A_\sigma,A_\tau]=&0\,,\\
\omega\,\left(\partial_{\bar{z}} A_\sigma -\partial_\sigma A_{\bar{z}}+[A_{\bar{z}},A_\sigma]
\right)=&0\,,\\
\omega\,\left(\partial_{\bar{z}} A_\tau -\partial_\tau A_{\bar{z}}+[A_{\bar{z}},A_\tau]
\right)=&0\,.
\end{align}
The factor $\omega$ is kept in order to cover the case $\partial_{\bar{z}}A_\sigma$ 
and $\partial_{\bar{z}}A_\tau$ are distributions on $\mathbb{C}P^1$ supported by $\mathfrak{z}$\,.

\medskip

It is also helpful to rewrite the boundary equation of motion (\ref{boundary eom})  into the form
\begin{align}
\sum_{x\in\mathfrak{p}}\sum_{p\geq0}\left(\operatorname{res}_x \xi_x^p \omega\right)\epsilon^{ij}\frac{1}{p!}\partial_{\xi_x}^p
\langle A_i,\delta A_j \rangle\big|_{\cM \times \{x\}}=0\,,\label{general boundary}
\end{align}
where $\epsilon^{ij}$ is the antisymmetric tensor.
Here the local holomorphic coordinates $\xi_x$ is defined as $\xi_x:=z-x$ 
for $x\in\mathfrak{p}\backslash\{\infty\}$ and $\xi_\infty:=1/z$ if $\mathfrak{p}$ 
includes the point at infinity. The relation (\ref{general boundary}) manifestly shows that the boundary equation 
of motion does not vanish only on $\cM\times\mathfrak{p}$\,.

\subsection*{Gauge invariance}

Let us discus here the gauge invariance of the action (\ref{4dcs})\,.  

\medskip

One may consider a transformation 
\begin{align}
A\mapsto A^u:=uAu^{-1}-duu^{-1}\,,  
\label{gauge transformation}
\end{align}
where $u$ is a $G^{\mathbb{C}}$-valued function defined on 
$\cM\times\mathbb{C}P^1$\,. 
Under this transformation, the field strength $F(A)$ transforms as
\begin{align}
F(A)\mapsto F(A^u)=uF(A)u^{-1}\,.
\end{align}

\medskip 

At the off-shell level,  the action (\ref{4dcs}) transforms under the transformation 
(\ref{gauge transformation}) as
\begin{align}
S[A^u]=S[A]-\frac{i}{4\pi}\int_{\cM\times \mathbb{C}P^1} \omega \wedge I_{WZ}[u]- \frac{i}{4\pi}\int_{\cM\times \mathbb{C}P^1} \omega\wedge d\langle u^{-1}du,A\rangle\,, 
\label{gauge variation}
\end{align}
where $I_{WZ}[u]$ is the Wess-Zumino (WZ) 3-form defined as
\begin{align}
I_{WZ}[u]:=\frac{1}{3}\langle u^{-1}du,u^{-1}du\wedge u^{-1}du\rangle\,.
\end{align}
Hence the action (\ref{4dcs}) is invariant if the gauge parameter $u$ satisfies
\begin{align}
\frac{i}{4\pi}\int_{\cM\times \mathbb{C}P^1} \omega \wedge I_{WZ}[u]=0\,,
\qquad u|_{\mathfrak{p}}=1\,. \label{u-cond}
\end{align}
Thus the transformation (\ref{gauge transformation}) can be regarded as a gauge transformation 
with $u$ satisfying the condition (\ref{u-cond})\,.

\medskip 

At the on-shell level, the bulk equation of motion (\ref{bulk eom}) is invariant 
under the gauge transformation, but the boundary equation 
of motion (\ref{boundary eom}) may vary in general. Hence one may consider two kinds of 
gauge transformation. 1) A transformation $A\mapsto A^u$ which preserves 
the boundary equation of motion is called  a ``gauge transformation'', and 
2) a general off-shell gauge transformation is called a ``formal gauge transformation''.

\subsection{Lax form}

Let us then introduce the Lax form. 

\medskip

Considering a formal gauge transformation
\begin{align}
A=-d\hat{g}\hat{g}^{-1}+\hat{g}\mathcal{L}\hat{g}^{-1}\,, 
\label{L def}
\end{align}
for a smooth function $\hat{g}:\cM\times\mathbb{C}P^1
\rightarrow G^{\mathbb{C}}$, 
we can always choose the gauge
\begin{align}
\mathcal{L}_{\bar{z}}=0\,,\label{gauge fix}
\end{align}
by taking $A_{\bar{z}}$ to be a pure-gauge component. 
Thus, the 1-form $\mathcal{L}$ takes the form
\begin{align}
\mathcal{L}:=\mathcal{L}_{\sigma}d\sigma+\mathcal{L}_{\tau}d\tau\,,
\end{align}
which is referred to as the Lax form. This will be regarded as a Lax pair for 2D theory 
in our later discussion. 

\medskip 

In terms of the Lax form $\mathcal{L}$, the bulk equations of motion are expressed as
\begin{align}
\partial_\tau \mathcal{L}_\sigma -\partial_\sigma \mathcal{L}_\tau+&[\mathcal{L}_\tau,
\mathcal{L}_\sigma]=0\,,\\
\omega\wedge \partial_{\bar{z}}\mathcal{L}&=0\,.\label{L holomorphic}
\end{align}
It follows that $\mathcal{L}$ is a meromorphic 1-form with poles at the zeros of $\omega$\,,  
namely $\mathfrak{z}$ is regarded as the set of poles of $\mathcal{L}$\,. 
Note here that because the transformation (\ref{L def}) is just a formal gauge transformation, 
the Lax form $\mathcal{L}$ does not satisfy the original boundary condition (\ref{boundary eom}).

\medskip

By substituting (\ref{L def}) into the action (\ref{4dcs}) and using the relation 
(\ref{gauge variation})\,, one can obtain the following expression: 
\begin{align}
S[A]=-\frac{i}{4\pi}\int_{\cM\times \mathbb{C}P^1} \omega \wedge I_{WZ}[\hat{g}]- \frac{i}{4\pi}\int_{\cM\times \mathbb{C}P^1} \omega\wedge d\langle 
\hat{g}^{-1}d\hat{g},\mathcal{L}\rangle\,. \label{action-4d}
\end{align}
This is still a 4D action. In order to get a 2D action from this expression, 
one needs to discuss a bit more as explained in the next subsection.

\subsection{From 4D to 2D via the archipelago conditions}

As explained in \cite{DLMV}, if $\hat{g}$ satisfies the {\it archipelago conditions}, 
which will be defined below, the action (\ref{action-4d}) can be reduced to 
the 2D action with the WZ term for each point in $\mathfrak{p}$\,.

\medskip

The archipelago conditions for 
$\hat{g}$ 
are defined as follows: 
\begin{description}
\item[] There exist open disks $V_x$, $U_x$ for each $x\in\mathfrak{p}$ 
such that $\{x\}\subset V_{x}\subset U_{x}$ and
\item[\hspace{6pt}i)] $U_x\cap U_y=\phi$ if $x\neq y$ for all $x,y\in \mathfrak{p}$,
\item[\hspace{3pt}ii)] $\hat{g}=1$ outside $\cM\times \cup_{x\in\mathfrak{p}}U_x$,
\item[iii)] $\hat{g}|_{\cM\times U_x}$ depends only on $\tau,\sigma$  
and the radial coordinate $|\xi_x|$ where $\xi_x$ is the local holomorphic coordinate,
\item[iv)] $\hat{g}|_{\cM\times V_x}$ depends only on $\tau$ and $\sigma$, that is, 
$g_{x}:=\hat{g}|_{\cM\times V_x}=\hat{g}|_{\cM\times \{x\}}$.
\end{description}
The conditions i) and ii) can always be satisfied thanks to a gauge transformation, 
but it is necessary to take an appropriate boundary condition so as to satisfy  iii) and iv)\,. 
The archipelago conditions say that $\mathcal{L}$ and $A$ are identical outside 
the ``islands'' $U_x$ but they may be different inside $U_x$\,. 
To be more precise, while the Lax form $\mathcal{L}$ is meromorphic in $\mathcal{M}\times \mathbb{C}P^1$ due to the bulk equation of motion (\ref{L holomorphic}), $A$ is modified to satisfy the boundary equation of motion (\ref{boundary eom}) by a formal gauge transformation in the islands $U_x$\,.

\medskip

If $\hat{g}$ satisfies the archipelago conditions, the 4D action (\ref{4dcs}) can be reduced into 
a 2D action with the WZ term by performing an integral 
over $\mathbb{C}P^1$. The resulting action is given by
\begin{align}
S\left[\left\{g_{x}\right\}_{x \in \mathfrak{p}}\right]=\frac{1}{2} \sum_{x \in \mathfrak{p}} \int_{\cM}\left\langle\operatorname{res}_{x} (\varphi \,\mathcal{L}), g_{x}^{-1} d g_{x}\right\rangle
+\frac{1}{2} \sum_{x \in \mathfrak{p}}\left(\operatorname{res}_{x} \omega\right) \int_{\cM\times[0,R_x]} I_{\mathrm{WZ}}\left[g_{x}\right]\,,\label{2d action}
\end{align}
where $R_x$ is the radius of the open disk $U_x$\,. Due to the archipelago conditions iii), 
only the integral with respect to the radial direction remains in the second term of (\ref{2d action}). 

\medskip

Note here that the action (\ref{2d action}) still has a 2D gauge invariance. 
Under the transformation with a gauge parameter $h:\cM\rightarrow G$\,,
\begin{align}
g_x\mapsto g_x h\,,\qquad \mathcal{L}\mapsto h^{-1}\mathcal{L}h+h^{-1}dh\,,\label{2d gauge}
\end{align}
 the action (\ref{2d action}) is indeed invariant. 
This can be seen as the residual gauge symmetry after taking 
the gauge fixing (\ref{gauge fix})\,.

\subsubsection*{Reality condition}
It is natural to impose some conditions for the forms of $\omega$ and the boundary conditions on $A$ 
so that the reality of the 4D action (\ref{4dcs}) and the resulting action (\ref{2d action}) 
is ensured \cite{DLMV}. 
For a complex coordinate $z$, complex conjugation $z\mapsto \cc{z}$ defines an involution $\mu_{\mathrm{t}}:\mathbb{C}P^{1}\to\mathbb{C}P^{1}$\,.
Let $\tau:\mathfrak{g}^{\mathbb{C}}\to\mathfrak{g}^{\mathbb{C}}$ be an anti-linear involution. 
Then the set of the fixed point under $\tau$ is a real Lie subalgebra $\mathfrak{g}$ of $\mathfrak{g}^{\mathbb{C}}$.
The anti-linear involution $\tau$ satisfies
\begin{align}
\cc{\langle B,C \rangle} = \langle \tau B, \tau C \rangle\,, \qquad ^{\forall}B,C\in \mathfrak{g}^{\mathbb{C}}\,.
\end{align}
The associated operation to the Lie group $G$ is denoted by $\tilde{\tau}: G^{\mathbb{C}}\to G^{\mathbb{C}}$\,.

\medskip

Introducing these involutions, one can see that the reality of the action (\ref{2d action}) is ensured by the conditions
\begin{align}
\cc{\omega}=&\mu_{\mathrm{t}}^{*}\omega\,,\label{reality omega}  \\
\tau A=&\mu_{\mathrm{t}}^{*}A\,.\label{reality A}  
\end{align}
Recalling the relation (\ref{L def}), we suppose that 
\begin{align}
 \tilde{\tau} \hat{g}=\mu_{\mathrm{t}}^{*}\hat{g}\,,\qquad & \tau \mathcal{L}=\mu_{\mathrm{t}}^{*}\mathcal{L}\,,\label{reality L}
\end{align}
so as to satisfy (\ref{reality A}).

\subsection{The $\eta$-deformed PCM from the rational description} \label{rational description}

It is instructive to explain how to derive the action of the $\eta$-deformed PCM 
and the associated Lax pair, as an example. 
We employ the rational description as in \cite{DLMV}, though the left and right symmetries 
of PCM are exchanged here. Namely, the left symmetry is broken in \cite{DLMV} 
while the right symmetry is broken in our discussion for the comparison with the results obtained in \cite{Matsumoto:2015jja,KMY-monod}.

\medskip

Let us begin with a twist function $\varphi$ given by \cite{KMY-cQA,KMY-monod}
\begin{align}
\omega=\varphi(\zL)\,d\zL=\frac{K}{1-c^2\eta^2}\frac{1-\zL^2}{\zL^2-c^2\eta^2}\,d\zL\,. 
\label{rational}
\end{align}
Here $K$ and $\eta$ are real constants. The value of $\eta$ measures the deformation. 
Then $c$ is a constant parameter characterizing the mCYBE, 
\begin{align}
[R(x),R(y)]-R\left([R(x),y]+[x,R(y)]\right)=-c^2[x,y] \qquad
(x,y\in\mathfrak{g},\; R\in \operatorname{End}\mathfrak{g})\,. 
\label{mCYBE}
\end{align}
There are two choices for the value of $c$\,, (i) $c=1$ (split type) 
and (ii) $c=i$ (non-split type). 
In particular, we are interested in a skew-symmetric $R$-operator satisfying
\begin{align}
\langle R(X),Y \rangle=-\langle X,R(Y) \rangle\,, \qquad \forall X,Y\in\mathfrak{g}\,.
\end{align}
 
For concreteness, we set $c=i$ in the following. Then the twist function (\ref{rational}) reads 
\begin{align}
\varphi(\zL)=\frac{K}{1+\eta^2}\frac{1-\zL^2}{\zL^2+\eta^2}\,.
\end{align}
Then the meromorphic 1-form $\omega$ has simple poles $\mathfrak{p}_1=\{\pm i\eta\}$\,, 
double pole $\mathfrak{p}_2=\{\infty\}$ and single zeros $\mathfrak{z}=\{\pm 1\}$\,. 
 
\medskip
 
Since the residues satisfy $\operatorname{res}_{-i\eta}\omega 
=\overline{\operatorname{res}_{+i\eta}\omega}$\,, 
the boundary equation of motion (\ref{general boundary}) can be expressed as
\begin{align}
 \epsilon^{ij}\llangle  A_{i}|_{+i\eta},\delta A_j|_{+i\eta} 
 \rrangle_{\mathfrak{g}^{\mathbb{C}};\pm i\eta}
 +(\operatorname{res}_{\infty}\omega)\epsilon^{ij}\langle A_i|_{\infty},\delta A_j|_{\infty} \rangle
  +(\operatorname{res}_{\infty}\xi_x\omega)\epsilon^{ij}\partial_{\xi_x}\langle A_i,\delta A_j 
\rangle|_{\infty} = 0\,. 
\label{rational boundary}
\end{align}
Here $\llangle\cdot,\cdot\rrangle_{\mathfrak{g}^{\mathbb{C}};\pm i\eta}:\mathfrak{g}^{\mathbb{C}}\times\mathfrak{g}^{\mathbb{C}}\rightarrow \mathbb{R}$ is the non-degenerate symmetric adjoint-invariant bilinear form defined as
 \begin{align}
 \llangle x,x'\rrangle_{\mathfrak{g}^{\mathbb{C}};
 \pm i\eta}:=2\operatorname{Re}\left((\operatorname{res}_{+i\eta}\omega) 
 \langle x,x'\rangle\right)\,,
\end{align}
for $x,x'\in\mathfrak{g}^{\mathbb{C}}$\,. 
 
\medskip
 
To solve the boundary equation of motion, we set the following conditions: 
\begin{align}
 A_i|_{+i\eta}&\in\mathfrak{g}_{R}\,,\label{A in gR}\\
  A_i|_{\infty}&=0\,,\label{A=0}
\end{align}
where $\mathfrak{g}_{R}:=\{(R-i)x|x\in\mathfrak{g}\}$. 
The first condition (\ref{A in gR}) ensures that the first term of (\ref{rational boundary}) vanishes, and the second condition (\ref{A=0}) deletes the second and third terms.
The boundary condition (\ref{A in gR}) means how to decompose the Lie algebra 
$\mathfrak{g}^{\mathbb{C}}=\mathfrak{g}_R\oplus\mathfrak{g}$. 
Here, the set $(\mathfrak{g}^{\mathbb{C}},\mathfrak{g}_R,\mathfrak{g})$ is a Manin triple, 
and the fact that $\mathfrak{g}_R$ is a Lie subalgebra of $\mathfrak{g}^{\mathbb{C}}$ 
plays a crucial role to satisfy the archipelago conditions iii) and iv) 
preserving the boundary equation of motion (\ref{rational boundary}).
 
 \medskip
 
The general discussion suggests that the Lax form $\mathcal{L}$ 
should have single poles at points in $\mathfrak{z}$\,. 
Hence it is natural to assume the following expression of $\mathcal{L}$: 
\begin{align}
 \mathcal{L}=\frac{V_{+}}{\zL+1}\,d\sigma^{+}+\frac{V_{-}}{\zL-1}\,d\sigma^{-} 
 +U_{+}\,d\sigma^{+}+U_{-}\,d\sigma^{-}\,. 
 \label{rational ansatz}
\end{align}
Here $\sigma^{\pm}$ are the light-cone coordinates on $\cM$ defined as 
\begin{eqnarray}
\sigma^{\pm}:= \frac{1}{2} (\tau\pm\sigma)\,, \qquad d\sigma^{+}\wedge d\sigma^{-}=-\frac{1}{2}d\tau\wedge d\sigma\,,
\end{eqnarray} 
and $V_{\pm},U_{\pm}:\cM\rightarrow \mathfrak{g}$ are smooth functions.
Using the 2D gauge symmetry and the freedom accompanied with the choice of the Lie subalgebra 
of $\mathfrak{g}^{\mathbb{C}}$ allow us to set an archipelago type field $\hat{g}$ like 
\begin{align}
\hat{g}_{\pm i\eta}=g^{-1}\,,\qquad \hat{g}_{\infty}=1\,.
\end{align}
Here $g:\cM\rightarrow G$ is a smooth function and the reality of $g$ is ensured by appropriate equivariant property of $\hat{g}$ as discussed in \cite{DLMV}. 

\medskip

Then the relation between $A$ and $\mathcal{L}$ (\ref{L def}) leads to 
\begin{align}
A|_{\pm i\eta}=g^{-1}dg+\operatorname{Ad}_{g^{-1}}\mathcal{L}|_{\pm i\eta}\,,\qquad A|_{\infty}=\mathcal{L}|_{\infty}\,,\label{rational AL}
\end{align}
where $\operatorname{Ad}_g$ is defined as 
\begin{equation}
\operatorname{Ad}_g x:=g\,x\,g^{-1}\,, \quad \mbox{for}~~x\in \mathfrak{g}\,. 
\end{equation} 
The boundary conditions (\ref{A in gR}) and (\ref{A=0}) can be expressed explicitly as
\begin{align}
(R+i)A_i|_{i\eta}=(R-i)A_i|_{-i\eta}\,,\qquad A|_{\infty}=0\,.\label{rational Ri}
\end{align}
Introducing the left-invariant 1-form $j:=g^{-1}dg$ and $R_g:=\operatorname{Ad}_{g}^{-1}\circ R\circ \operatorname{Ad}_{g}$, the unknown functions $V_{\pm}$ and $U_{\pm}$ are determined by combining (\ref{rational ansatz}), (\ref{rational AL}) and (\ref{rational Ri}) as
\begin{align}
V_{\pm}=\mp\frac{\eta^2+1}{1\mp\eta R_{g^{-1}}}\partial_{\pm}gg^{-1}=\mp g\left(\frac{\eta^2+1}{1\mp\eta R}\,j_{\pm}\right)g^{-1}\,, \qquad U_{\pm}=0\,. 
\end{align}
As a result, the Lax form is given by  
\begin{align}
\mathcal{L}=g\left[-\left(\frac{\eta^2+1}{1-\eta R}\right)\frac{1}{\zL+1}\, j_{+}\, d\sigma^{+}+\left(\frac{\eta^2+1}{1+\eta R}\right)\frac{1}{\zL-1}\,j_{-}\,d\sigma^{-}\right]g^{-1}\,. 
\label{rational Lax}
\end{align}

\medskip 

Finally, let us evaluate  the 2D action (\ref{2d action})\,. 
The residues $\operatorname{res}_{\pm i \eta}(\varphi \mathcal{L})$ are computed as 
\begin{align}
\operatorname{res}_{\pm i \eta}(\varphi \mathcal{L})
=\mp\frac{K}{2i\eta}g\left(\frac{1\mp i\eta}{1-\eta R}\,j_{+}d\sigma^{+}+\frac{1\pm i\eta}{1+\eta R}\,j_{-}d\sigma^{-}\right)g^{-1}\,,
\end{align}
and $\hat{g}^{-1}d\hat{g}|_{\infty}=0$\,. Hence the resulting action is given by 
\begin{align}
S[g]
=&\frac{K}{2}\int_{\cM}d\tau\wedge d\sigma\left\langle j_{-},\frac{1}{1-\eta R}\,j_{+}\right\rangle\,.\label{rational action}
\end{align}
The action (\ref{rational action}) and the Lax form (\ref{rational Lax}) are equivalent to 
the ones of the $\eta$-deformed PCM \cite{Klimcik1,Klimcik2}.

\section{From the trigonometric description}\label{trigonometric description} 

In this section, we will start from a twist function in the {\it trigonometric} description 
and try to reproduce the $\eta$-deformed PCM. 
Then the spectral parameter $z_{\rm R}$ takes a value on a cylinder rather than $\mathbb{C}P^1$ 
parametrized by $z_{\rm L}$\,. The cylinder geometry is a characteristic of the trigonometric class of 
integrable system and $z_{\rm R}$ should be distinguished from $z_{\rm L}$ \cite{KMY-monod} 
(though they are related each other via a certain relation, as will be explained later). 

\medskip 

In fact, the cylinder is equivalent to a couple of $\mathbb{C}P^1$'s. When the rational description  
is adopted, it is necessary to take account of the gauge transformed monodromy separately so as to see the equivalence. 
But by starting from the trigonometric description, 
one can discuss the whole space of spectral parameter by construction all at once.  

\medskip 

In the analysis here, by starting from the trigonometric description, we could not only reproduce 
the well-known result, but also discover a new type of 
YB deformation as a byproduct. This is the main result of our paper.

\subsection{Twist function}

The meromorphic 1-form $\omega$ in the trigonometric description 
\cite{KMY-cQA,KMY-monod} is given by
\begin{align}
\omega=\frac{\sinh (\alpha-\zR) \sinh (\alpha+\zR)}{\sinh{\alpha} 
\cosh{\alpha} \,\sinh ^{2} \zR} \, d \zR = \varphi_{\mathsf{c}}(\zR) \, d\zR\,,
\label{eq:twist-eta-z}
\end{align}
where $\alpha$ is a pure imaginary parameter. 
Since $\varphi_{\mathsf{c}}(\zR)$ has the following periodicity: 
\begin{align}
\varphi_{\mathsf{c}}(\zR)=\varphi_{\mathsf{c}}(\zR+2\pi i)\,, 
\end{align}
the fundamental region of $z_{\rm R}$ can be taken as  
\begin{equation}
\mathbb{C}/\mathbb{Z}=\left\{\zR\in\mathbb{C}\,\bigg| -\frac{\pi}{2}<\operatorname{Im}\zR< \frac{3\pi}{2} \right\}\,. 
\end{equation}
This cylinder $\mathbb{C}/\mathbb{Z}$ can be mapped to a plane 
$\mathbb{C}^{\times}:=\mathbb{C}\backslash \{0\}$ 
via the map 
\begin{equation}
\wR:=\exp\zR\,. 
\end{equation}
Then $\omega$ in (\ref{eq:twist-eta-z}) is rewritten as 
\begin{align}
\omega&=\frac{4\left(e^{2 \alpha}-\wR^{2}\right)\left(e^{2 \alpha} \wR^{2}-1\right)}{\left(e^{4 \alpha}-1\right) \wR\left(\wR^{2}-1\right)^{2}} \, d\wR
=\varphi(\wR) \, d\wR\,.
\label{eq:twist-eta-w}
\end{align}
In the following, we take $\wR$ as the global holomorphic coordinate.

\subsubsection*{Zeros and poles of $\omega$}

The meromorphic 1-form $\omega$ in (\ref{eq:twist-eta-w}) has single zeros
\begin{align}
\mathfrak{z}=\{e^{-\alpha},\mathrm{e}^{\alpha},-e^{-\alpha},-e^{\alpha}\}\,,
\end{align}
and the set of single poles $\mathfrak{p}_1$ and double poles $\mathfrak{p}_2$
\begin{align}
\mathfrak{p}_1=\{0,\infty\}\,, \qquad \mathfrak{p}_2=\{-1,+1\}\,,\qquad (\mathfrak{p}=\mathfrak{p}_1\cup\mathfrak{p}_2)\,.
\end{align} 
The location of the zeros and poles is shown in Fig.\ 1.
The residues at the poles are given by
\begin{align}
&\operatorname{res}_{\wR=1}\omega=0\,,  
&&\hspace*{-1.5cm} \operatorname{res}_{\wR=1} \xi_{1}\,\omega =\tanh{\alpha}\,, \notag \\
&\operatorname{res}_{\wR=-1}\omega=0\,,& & \hspace*{-1.5cm}
\operatorname{res}_{\wR=-1} \xi_{-1}\,\omega 
=-\tanh{\alpha}\,,  \\ 
&\operatorname{res}_{\wR=0}\omega =-\frac{2}{\sinh{2\alpha}}\,, & & \hspace*{-1.5cm}
\operatorname{res}_{\wR=\infty}\omega=+\frac{2}{\sinh{2\alpha}}\,, \notag
\end{align}
where the local holomorphic coordinates are defined by $\xi_{\pm1}:=\wR \mp1$.

\begin{figure}[ht]
\vspace*{1cm}
\begin{center}
\includegraphics[width=13.0cm]{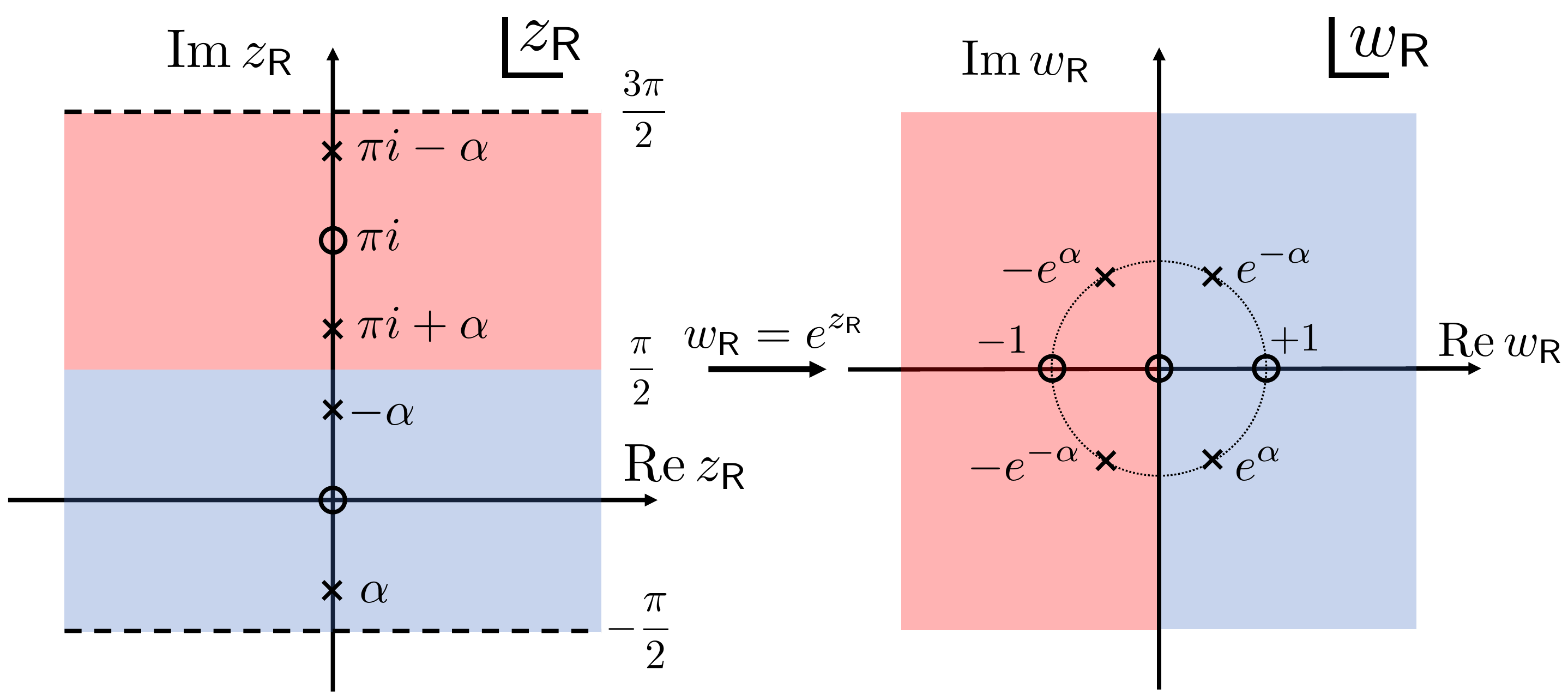}
\caption{\footnotesize $\mathbb{C}/\mathbb{Z}$ (with $z_{\rm R}$) 
and $\mathbb{C}^{\times}$ (with $w_{\rm R}$)\,. 
The poles and zeros of $\omega$ are denoted by $\circ$ and $\times$\,, respectively. }
\end{center}\label{figure1}
\end{figure}

\subsubsection*{The reality condition of $\omega$}

Let us consider the reality condition in the trigonometric description. 
The complex plane of spectral parameter is different from the rational description. 
It is necessary to discuss a possible involution in more detail\footnote{The authors greatly 
appreciate the referee's comment and suggestion for this point.}. 

\medskip 

An involution $\mu_{\mathrm{t}}$ may be defined with complex conjugation for 
$\zR':=i\zR$ as follows:
\begin{align}
\mu_{\mathrm{t}}:
\zR' \to \overline{\zR'}
\quad\Longleftrightarrow\quad \zR \to -\overline{\zR}
\quad\Longleftrightarrow\quad \wR \to \frac{1}{\overline{\wR}}\,.
\end{align}
Note that the points $\wR=\pm 1$ are fixed points of $\mu_{\mathrm{t}}$\,.
Then by using this $\mu_{\rm t}$ the reality condition (\ref{reality omega}) is realized as 
\begin{align}\begin{split}
\cc{\omega}=&\cc{\left(\frac{\sin (i\alpha-\zR') \sin (i\alpha+\zR')}{\sin{i\alpha} \cos{i\alpha} \,\sin^{2} \zR'} \, d\zR'\right)}
=\frac{\sin (i\alpha-\cc{\zR'}) \sin (i\alpha+\cc{\zR'})}{\sin{i\alpha} \cos{i\alpha} \,\sin^{2} \cc{\zR'}} \, d\cc{\zR'}\\
=&\mu_{\mathrm{t}}^{*}\omega\,.
\end{split}\end{align}

\subsection{The boundary condition}

In the 4D CS theory (\ref{4dcs}), the base space $\mathcal{M} \times \mathbb{C}P^1$ 
is replaced by $\mathcal{M} \times \mathbb{C}^{\times}$\,. However, 
the bulk and boundary equations of motion in (\ref{bulk eom}) and (\ref{boundary eom}) 
remain unchanged. 

\medskip  

For $\omega$ in (\ref{eq:twist-eta-w})\,,  
the boundary equation of motion (\ref{general boundary}) is expressed as 
\begin{align}
0=&\quad\left(\operatorname{res}_{\wR=1}\omega\right)\epsilon^{ij}\left\langle A_i|_{1},\delta A_j|_{1}\right\rangle+\left(\operatorname{res}_{\wR=1}\xi_{1}\omega\right)\epsilon^{ij}\partial_{\xi_{1}}\left\langle A_i|_{1},\delta A_j|_{1}\right\rangle\nonumber \\
&+\left(\operatorname{res}_{\wR=-1}\omega\right)\epsilon^{ij}\left\langle A_i|_{-1},\delta A_j|_{-1}\right\rangle+\left(\operatorname{res}_{\wR=-1}\xi_{-1}\omega\right)\epsilon^{ij}\partial_{\xi_{-1}}\left\langle A_i|_{-1},\delta A_j|_{-1}\right\rangle\nonumber\\
&+\left(\operatorname{res}_{\wR=0}\omega\right)\epsilon^{ij}\left\langle A_i|_{0},\delta A_j|_{0}\right\rangle+\left(\operatorname{res}_{\wR=\infty}\omega\right)\epsilon^{ij}\left\langle A_i|_{\infty},\delta A_j|_{\infty}\right\rangle\,.
\label{eq:bEOM-mCYBE-nonsp}
\end{align}
Then the sufficient conditions are given by 
\begin{align}
\epsilon^{ij}\llangle\left(A_i|_1,(\partial_{\xi_1}A_i)|_{1}\right),&\delta \left(A_j|_1,(\partial_{\xi_1}A_j)|_{1}\right) \rrangle_{\mathfrak{t};1}=0\,  \label{bc-w1}\\
\epsilon^{ij}\llangle\left(A_i|_{-1},(\partial_{\xi_{-1}}A_i)|_{-1}\right),&\delta \left(A_j|_{-1},(\partial_{\xi_{-1}}A_j)|_{-1}\right) \rrangle_{\mathfrak{t};-1}=0\,,\label{bc-w-1}\\
\epsilon^{ij}\llangle (A_i|_{0},A_i|_{\infty}),&\delta(A_j|_{0}, A_j|_{\infty})\rrangle_{\mathfrak{h};\wR=0,\infty}=0\,,\label{eq:bEOM-mCYBE}
\end{align}
where the above bilinear forms are defined as, respectively, 
\begin{align}
\llangle(x,y),(x',y')\rrangle_{\mathfrak{t};1}&:=(\operatorname{res}_{\wR=1}\omega)\langle x,x'\rangle+(\operatorname{res}_{\wR=1}\xi_{1}\omega)(\langle x,y'\rangle+\langle x',y\rangle) \nonumber \\
&=\tanh{\alpha}(\langle x,y'\rangle+\langle x',y\rangle)\,,\\
\llangle(x,y),(x',y')\rrangle_{\mathfrak{t};-1}&:=(\operatorname{res}_{\wR=-1}\omega)\langle x,x'\rangle+(\operatorname{res}_{\wR=-1}\xi_{-1}\omega)(\langle x,y'\rangle+\langle x',y\rangle) \nonumber \\
&=-\tanh{\alpha}(\langle x,y'\rangle+\langle x',y\rangle)\,,\\
\llangle (x,y),(x',y') \rrangle_{\mathfrak{h};\wR=0,\infty}&:=(\operatorname{res}_{\wR=0}\omega)\langle x,x'\rangle+(\operatorname{res}_{\wR=\infty}\omega)\langle y,y'\rangle   \nonumber \\
&=-\frac{2}{\sinh 2\alpha} 
(\langle x,x' \rangle - \langle y,y' \rangle )\,. \label{eq:d-inner}
\end{align}

To satisfy these boundary conditions, we assign Drinfeld doubles to the bilinear forms as follows: 
\begin{align}
\mathfrak{t}=(\mathfrak{g}\ltimes\{0\})\oplus (\{0\}\ltimes\mathfrak{g}_{\mathrm{ab}})
\end{align}
for $\wR=1,-1$\,, and a new Drinfeld double
\begin{align}
\mathfrak{h}:=
\mathfrak{g}^\delta\oplus\mathfrak{g}_{R}
\label{def:d=gd+gR}
\end{align}
for $\wR=0,\infty$\,. Here $\mathfrak{g}_\mathrm{ab}$ is an abelian copy of $\mathfrak{g}$\,,
 $\mathfrak{g}^\delta$ and $\mathfrak{g}_R$ are defined as 
\begin{align}
\mathfrak{g}_R&:= \left\{((R-i)x,(R+i)x)|x\in\mathfrak{g}\right\}\,,\label{def:gR}\\
\mathfrak{g}^\delta&:=\left\{(x,x)|x\in\mathfrak{g}\right\}\,.\label{def:gd}
\end{align}
As a result, $A_{i}$ is supposed to satisfy
\begin{alignat}{2}
&(A_i|_{\wR=1},(\partial_{\xi_1} A_{i})|_{\wR=1})&&\in\{0\}\ltimes \mathfrak{g}_{\mathrm{ab}}\,,
\label{manin1}\\
&(A_i|_{\wR=-1},(\partial_{\xi_{-1}} A_{i})|_{\wR=-1})&&\in\{0\}\ltimes \mathfrak{g}_{\mathrm{ab}}\,,\label{manin-1}\\
&(A_{i}|_{\wR=0},A_{i}|_{\wR=\infty})&& \in \mathfrak{g}_{R}\,,
\label{manin0}
\end{alignat}
by taking account of the reality condition (\ref{reality A}).
For a short review of Manin triple and Drinfeld double, see Appendix \ref{Appendix drinfeld}.

\medskip

These choices indeed ensure that the boundary conditions are satisfied, 
because $\{0\}\ltimes \mathfrak{g}_{\mathrm{ab}}$ and $\mathfrak{g}_{R}$ 
are isotropic with respect to the bilinear forms defined above, that is,
\begin{align}
&\llangle(0,y),(0,y')\rrangle_{\mathfrak{t};1}=\tanh{\alpha}(\langle 0,y'\rangle+\langle 0,y\rangle)=0\,,\\
&\llangle(0,y),(0,y')\rrangle_{\mathfrak{t};-1}=-\tanh{\alpha}(\langle 0,y'\rangle+\langle 0,y\rangle)=0\,,\\
&\llangle ((R-i)x,(R+i)x),((R-i)x',(R+i)x') \rrangle_{\mathfrak{h};\wR=0,\infty}\nonumber\\
=&-\frac{2}{\sinh{2\alpha}}\langle (R-i)x,(R-i)x'\rangle+\frac{2}{\sinh{2\alpha}}\langle (R+i)x,(R+i)x'\rangle\nonumber\\
=&-\frac{2}{\sinh{2\alpha}}\left(\langle Rx,Rx'\rangle-\langle x,x'\rangle\right) 
+ \frac{2}{\sinh{2\alpha}}\left(\langle Rx,Rx'\rangle-\langle x,x'\rangle\right) 
\nonumber\\
=&0\,.\label{3.23}
\end{align}
Note here that the skew-symmetricity of $R$ has been utilized in (\ref{3.23})\,. 

\medskip

In addition, these subspaces are found to be Lie subalgebras of $\mathfrak{t}$ and 
$\mathfrak{h}$\,, respectively. 
Thus these conditions are sufficient to derive 2D action.

\subsection{Lax form and 2D action}

Next, let us determine the associated Lax form and 2D action.  

\medskip 

The first is to determine the Lax form $\mathcal{L}$\,. The meromorphic 1-form $\omega$ 
has four single zeros in $\mathfrak{z}$\,, hence the Lax form should have simple poles 
at $\mathfrak{z}$\,. Assume that the light-cone components of  
$\mathcal{L}$ are represented by  
\begin{align}
\mathcal{L}_{+}&=\frac{V_{+}\, \wR+V_{+}'}{e^{2\alpha}\,\wR^2-1}+U_{+}\,,\label{trig Lax plus} \\
\mathcal{L}_{-}&=\frac{V_{-}\, \wR+V_{-}'}{\wR^2-e^{2\alpha}}+U_{-}\,,\label{trig Lax minus}
\end{align}
where $V_{\pm},V_{\pm}' ,U_{\pm}:\cM\rightarrow\mathfrak{g}^\mathbb{C}$ are smooth functions. 
Note here that the reality condition (\ref{reality L}) is now realized with $\mu_{\rm t}$ 
as follows: 
\begin{eqnarray}
\tau{\cL_{\pm}}=\mu_{\mathrm{t}}^{*}\cL_{\pm}\,, \qquad \tilde{\tau} \hat{g}=\mu_{\mathrm{t}}^{*}\hat{g}\,. 
\end{eqnarray}

\medskip

Taking boundary conditions as (\ref{manin1}), (\ref{manin-1}) and (\ref{manin0}) means that
\begin{align}
A_{\pm}|_{\wR=+1}=0\,,\quad & A_{\pm}|_{\wR=-1}=0\,,\label{A=0 trig}\\
(R+i)A_{\pm}|_{\wR=0}=&(R-i)A_{\pm}|_{\wR=\infty}\,.
\end{align}
Thus, by using the relation (\ref{L def})\,, one can obtain the following expressions: 
\begin{align}
\hat{g}^{-1}\partial_{\pm}\hat{g}|_{\wR=+1}&=\pm\frac{V_{\pm}+V_{\pm}'}{e^{2\alpha}-1}+U_{\pm}\label{bccal1}\,,\\
\hat{g}^{-1}\partial_{\pm}\hat{g}|_{\wR=-1}&=\pm\frac{-V_{\pm}+V_{\pm}'}{e^{2\alpha}-1}+U_{\pm}\,,\\
(R+i)\left[\hat{g}(-\hat{g}^{-1}\partial_{+}\hat{g}-V_{+}'+U_{+})\hat{g}^{-1}\right]\big|_{\wR=0}
&=(R-i)\left[\hat{g}(-\hat{g}^{-1}\partial_{+}\hat{g}+U_{+})\hat{g}^{-1}\right]\big|_{\wR=\infty}\,,\\
(R+i)\left[\hat{g}(-\hat{g}^{-1}\partial_{-}\hat{g}-e^{-2\alpha}V_{-}'+U_{-})\hat{g}^{-1}\right]\big|_{\wR=0}
&=(R-i)\left[\hat{g}(-\hat{g}^{-1}\partial_{-}\hat{g}+U_{-})\hat{g}^{-1}\right]\big|_{\wR=\infty}\,.\label{bccal4}
\end{align}
Since the choice of the Drinfeld double (\ref{def:d=gd+gR}) enable us to take $\hat{g}|_{\wR=0}\in G$ (not $G^{\mathbb{C}}$), one can take $\hat{g}|_{\wR=0}=1$ by using  a 2D gauge invariance $g\mapsto gh\:(h\in G)$\,.
Furthermore, the condition $\tilde{\tau} \hat{g}=\mu_{\mathrm{t}}^{*} \hat{g}$ indicates $\tilde{\tau}(\hat{g}|_{\wR=0})=\mu_{\mathrm{t}}^{*}(\hat{g}|_{\wR=0})=\hat{g}|_{\wR=\infty}$\,, and then
\begin{align}
\hat{g}|_{\wR=0}=\hat{g}|_{\wR=\infty}=1\,. 
\end{align}
For the other poles, one may introduce two degrees of freedom $g$\,, $\tilde{g} \in G$ (not $G^{\mathbb{C}}$) like
\begin{align}
g:=\hat{g}|_{\wR=+1}\,,\qquad \tilde{g}:=\hat{g}|_{\wR=-1}\,,
\end{align}
because $\wR=\pm 1$ are fixed points of $\mu_{\mathrm{t}}$\,.

\medskip 

Then for $g$ and $\tilde{g}$\,, the left-invariant 1-forms can also be introduced as 
\begin{align}
j_{\pm}:=g^{-1}\partial_{\pm}g\,,\qquad
\tilde{j}_{\pm}:=\tilde{g}^{-1}\partial_{\pm}\tilde{g}\,.
\end{align}
Thus the boundary conditions (\ref{bccal1})-(\ref{bccal4}) lead to 
\begin{align}
\pm\frac{V_{\pm}+V_{\pm}'}{e^{2\alpha}-1}+U_{\pm}&=j_{\pm}\,,\\
\pm\frac{-V_{\pm}+V_{\pm}'}{e^{2\alpha}-1}+U_{\pm}&=\tilde{j}_{\pm}\,,\\
(R+i)(-V_{+}'+U_{+})&=(R-i)(+U_{+})\,,\\
(R+i)(-e^{-2\alpha}V_{-}'+U_{-})&=(R-i)(+U_{-})\,.
\end{align}
These relations can be solved by
\begin{align}
V_{\pm}&=\pm(e^{2\alpha}-1)\frac{j_\pm-\tilde{j}_\pm}{2}\,,\\
V_{+}^{'}&=\frac{1}{\coth{\alpha}-iR}(j_{+}+\tilde{j}_{+})\,,\\
V_{-}^{'}&=-\frac{e^{2\alpha}}{\coth{\alpha}+iR}(j_{-}+\tilde{j}_{-})\,,\\
U_{\pm}&=\pm\frac{1\mp\coth{\alpha}}{2(\coth{\alpha}\mp iR)}(j_{\pm}+\tilde{j}_{\pm})+\frac{1}{2}(j_{\pm}+\tilde{j}_{\pm})\,.
\end{align}
Thus the components of $\mathcal{L}$ in (\ref{trig Lax plus}) and (\ref{trig Lax minus}) are given by
\begin{align}
\mathcal{L}_{+}
=&\frac{\sinh{\alpha}}{2\sinh(\zR+\alpha)}(j_{+}-\tilde{j}_{+})+\frac{1}{2}\frac{i\coth(\zR+\alpha)+R}{i\coth{\alpha}+R}(j_{+}+\tilde{j}_{+})\,,
\label{Lp} 
\\
\mathcal{L}_{-}
=&-\frac{\sinh{\alpha}}{2\sinh(\zR-\alpha)}(j_{-}-\tilde{j}_{-})+\frac{1}{2}\frac{i\coth(\zR-\alpha)+R}{-i\coth{\alpha}+R}(j_{-}+\tilde{j}_{-})\,.
\label{Lm}
\end{align}
Then it is possible to put together these expressions into a single form like 
\begin{align}
\mathcal{L}_{\pm}=&
\pm\frac{\sinh{\alpha}}{2\sinh(\zR\pm\alpha)}(j_{\pm}-\tilde{j}_{\pm})+\frac{1}{2}\frac{i\coth(\zR\pm\alpha)+R}{\pm i\coth{\alpha}+R}(j_{\pm}+\tilde{j}_{\pm})\label{S3Lax}\\
=&\frac{1}{2}\frac{\sqrt{1+\eta^2(\frac{\tanh{\zR}}{\tanh{\alpha}})^2}}{1\pm\frac{\tanh{\zR}}{\tanh{\alpha}}}(j_{\pm}-\tilde{j}_{\pm})+
\frac{1}{2}\frac{1}{1\pm\frac{\tanh{\zR}}{\tanh{\alpha}}}\left(1\mp \frac{\eta\frac{\tanh{\zR}}{\tanh{\alpha}}(\eta\pm R)}{1\mp\eta R}\right)(j_{\pm}+\tilde{j}_{\pm})\,,
\label{eq:biLax}
\end{align}
where a deformation parameter $\eta$ has been introduced as
\begin{align}
-i\eta:=\tanh{\alpha}\,.
\end{align}
One can see that the Lax pair (\ref{S3Lax}) indeed satisfies the reality condition:
\begin{align}
\tau\mathcal{L}=&
\left[\pm\frac{\sinh{\alpha}}{2\sinh(-\cc{\zR}\pm\alpha)}(j_{\pm}-\tilde{j}_{\pm})+\frac{1}{2}\frac{i\coth(-\cc{\zR}\pm\alpha)+R}{\pm i\coth{\alpha}+R}(j_{\pm}+\tilde{j}_{\pm})\right]d\sigma^{\pm}\no\\
=&\mathcal{L}_{\pm}(-\cc{\zR})d\sigma^{\pm}\no\\
=&\mu_{\mathrm{t}}^{*}\mathcal{L}\,.
\end{align}
The last expression of $\mathcal{L}$ (\ref{eq:biLax}) will be useful for our later discussion. 

\medskip 

Next, let us derive the deformed action by evaluating the master formula (\ref{2d action})\,. 
The residues necessary for the derivation are computed as 
\begin{align}
\operatorname{res}_{\wR=1} \omega\wedge\mathcal{L}=&d\wR\wedge d\sigma^{+}\left[
-\frac{1}{2}(j_{+}-\tilde{j}_{+})-\frac{2e^{2\alpha}}{e^{4\alpha}-1}\frac{1}{\coth{\alpha}-iR}(j_{+}+\tilde{j}_{+})
\right]\nonumber\\
&+d\wR\wedge d\sigma^{-}\left[
\frac{1}{2}(j_{-}-\tilde{j}_{-})+\frac{2e^{2\alpha}}{e^{4\alpha}-1}\frac{1}{\coth{\alpha}+iR}(j_{-}+\tilde{j}_{-})
\right]\,,\\
\operatorname{res}_{\wR=-1} \omega\wedge\mathcal{L}=&d\wR\wedge d\sigma^{+}\left[
\frac{1}{2}(j_{+}-\tilde{j}_{+})-\frac{2e^{2\alpha}}{e^{4\alpha}-1}\frac{1}{\coth{\alpha}-iR}(j_{+}+\tilde{j}_{+})
\right]\nonumber\\
&+d\wR\wedge d\sigma^{-}\left[
-\frac{1}{2}(j_{-}-\tilde{j}_{-})+\frac{2e^{2\alpha}}{e^{4\alpha}-1}\frac{1}{\coth{\alpha}+iR}(j_{-}+\tilde{j}_{-})
\right]\,.
\end{align}
Note that for these residues, the contributions from $U_\pm$ vanish.\\
Using the fact
\begin{align}
\operatorname{res}_{\wR=\pm1}\omega=0\,,
\end{align}
the resulting 2D action is given by 
\begin{align}
S[\{ j,\tilde{j}\}]
=\frac{1}{4}\int_{\cM}&d\tau \wedge d\sigma\Biggl[\big\langle j_{+}-\tilde{j}_{+} , j_{-}-\tilde{j}_{-}\big\rangle
+\frac{2e^{2\alpha}}{e^{4\alpha}-1}\big\langle j_{+}+\tilde{j}_{+},\frac{2}{\coth{\alpha}+iR}(j_{-}+\tilde{j}_{-})\big\rangle\Biggr]\,.
\end{align}
In terms of the deformation parameter $\eta$\,,
the action can be expressed as
\begin{align}
S[\{ j,\tilde{j}\}]
=&\frac{1}{4}\int_{\cM}d\tau \wedge d\sigma\Bigl[
\big\langle j_{+}-\tilde{j}_{+},j_{-}-\tilde{j}_{-}\big\rangle
  +(1+\eta^2)\big\langle j_{+}+\tilde{j}_{+},\frac{1}{1+\eta R}(j_{-}+\tilde{j}_{-})\big\rangle\Bigr]\,.
\label{biaction}
\end{align}

\subsection{Relating $\tilde{j}$ to $j$}\label{relating j}

So far, the resulting Lax form and 2D action are written in terms of $j$ and $\tilde{j}$\,. 
In order to reproduce the well-known results, one needs to impose some relation 
between $j$ and $\tilde{j}$ like $\tilde{j} = f(j)$ so as to remove $\tilde{j}$\,. 
It seems likely that there should be some choices for the relation. 

\medskip

Note firstly that the action (\ref{biaction}) is invariant under exchange of $j$ and $\tilde{j}$\,.
This fact requires that any relation $\tilde{j}=f(j)$ must exhibit the $\mathbb{Z}_2$-grading 
property $f(f(j))=j$\,. In the following, we start with $j$ and discuss for the relation to $\tilde{j}$\,. 
Then, in order for $\tilde{j}$ to satisfy the bulk and boundary equations of motion (\ref{bulk eom}) 
and (\ref{boundary eom}), the flatness condition for $\tilde{j}$\,,  
\begin{align}
\partial_{+}\tilde{j}_{-}-\partial_{-}\tilde{j}_{+}+[\tilde{j}_{+},\tilde{j}_{-}]=0
\label{eq:flatness-j}
\end{align}
must be satisfied.  In summary, the relation $\tilde{j} = f(j)$ 
has to satisfy the $\mathbb{Z}_2$-grading $f\circ f =1$ and 
preserve the flatness condition for $\tilde{j}$\,. 

\medskip 

A trivial relation\footnote{One might think of that the case $\tilde{j}=-j$ should be possible. 
However, it is not the case because $j$ and $\tilde{j}$ do not satisfy the same flatness condition.} 
is given by 
\begin{align}
\text{i)}\qquad \tilde{j}=j\,.
\label{eqe:j=tj}
\end{align}
It works for any Lie algebra $\mathfrak{g}$ and reproduces the action (\ref{rational action}) 
of the $\eta$-deformed PCM, as we will see later.

\medskip

One may consider another choice of $\tilde{j}$ 
if $\mathfrak{g}$ is supposed to be a simple Lie algebra. 
The commutation relations in the standard Cartan form are given by 
\begin{align}
&[H_a,H_b]=0\,, \qquad (a,b=1,2,\dots,r)\,,\no\\
&[H_{a},E_{\alpha}]=\alpha(H_a)E_{\alpha}\,,\qquad [E_{\alpha},E_{-\alpha}]=\alpha(H_a)H_{a}\,,\no\\
 &[E_{\alpha}, E_{\beta}]=N_{\alpha,\beta}E_{\alpha+\beta}\qquad (\beta\neq-\alpha)\,,
\label{eq:comm}
\end{align}
where $H_a$ are the elements of the Cartan subalgebra of $\mathfrak{g}$\,.
An important point is that the commutation relations (\ref{eq:comm}) are invariant 
under the following transformation:\footnote{A similar exponential map has been 
discussed in a different context \cite{Beisert-exponential}.} 
\begin{align}
E_{\pm\alpha}\mapsto \exp \left(\lambda_{\pm\alpha}\right) E_{\pm\alpha}\,,\qquad 
H_{a}\mapsto H_{a} \qquad (\lambda_{\pm\alpha}\in\mathbb{C})\,,
\label{eq:generator-map-expR}
\end{align}
where the parameter $\lambda_{\alpha}$ satisfies 
\begin{align}
\lambda_{\alpha}+\lambda_{\beta}=\lambda_{\alpha+\beta}\,, \qquad \lambda_{\alpha}+\lambda_{-\alpha}=0\,.\label{eq:homomorphism}
\end{align}
We choose $\lambda_{\alpha}$ as
\begin{equation}
\lambda_{\alpha}=-ik_{\alpha}\pi\qquad (k_{\alpha}\in\mathbb{Z})\,.\label{integer}
\end{equation} 
There may be some manners to realize 
a map $\lambda:\mathbb{C}^r\rightarrow \mathbb{C}$ 
such that (\ref{eq:homomorphism}) and (\ref{integer}) are satisfied. 
For possible representations, see Appendix \ref{appendix:exp}.

\medskip 

By employing the symmetry (\ref{eq:generator-map-expR}), we can consider the second configuration
\begin{align}
\text{ii)}\qquad \tilde{j}=\exp(\pi \Sigma)\,j\,,
\label{eqe:tj=expRj}
\end{align}
where $\Sigma: \mathfrak{g}^{\mathbb{C}}\to \mathfrak{g}^{\mathbb{C}}$ is defined as 
\begin{align}
\Sigma(E_{\alpha}) =-i k_{\alpha}E_{\alpha}\,,\qquad 
\Sigma(H_a)=0\,. 
\end{align}
The exponential map (\ref{eqe:tj=expRj}) corresponds to the choice (\ref{integer}) in (\ref{eq:generator-map-expR}).
In this case, $\tilde{j}$ satisfies the same flatness condition (\ref{eq:flatness-j}) as $j$\,, and satisfies 
the $\mathbb{Z}_2$-property: 
\begin{equation} 
j= \exp(\pi\Sigma) \circ \exp(\pi\Sigma)\,j\,. 
\end{equation} 
Hence one may take this configuration. Indeed, 
the commutation relations in (\ref{eq:comm}) imply that the Lie algebra has a $\mathbb{Z}_2$-grading by assigning the grade 0 and 1 for the spaces where $k_{\alpha}$ is even and odd, respectively.

\subsubsection*{i)  Solution with $\tilde{j}=j$}
Let us first consider the configuration $\tilde{j}=j$\,.
 In this case, the Lax pair (\ref{Lp}), (\ref{Lm}) becomes
\begin{align}
\begin{split}
\cL_{+}^{\sfR}(\wR)=&+\frac{2}{e^{2\alpha}\wR^2-1}\frac{1}{\coth{\alpha}-iR}j_{+}+\frac{1-\coth{\alpha}}{\coth{\alpha}-iR}j_{+}+j_{+}
\,,\\
\cL_{-}^{\sfR}(\wR)=&-\frac{2}{e^{-2\alpha}\wR^2-1}\frac{1}{\coth{\alpha}+iR}j_{-}-\frac{1+\coth{\alpha}}{\coth{\alpha}+iR}j_{-}+j_{-}\,.
\label{eq:Lax-eta-R-w}
\end{split}
\end{align}
In terms of $\zR$\,, the Lax pair (\ref{eq:Lax-eta-R-w}) can be rewritten as
\begin{align}
\cL_{\pm}^{\sfR}(\zR)=&\frac{i\coth(\zR\pm\alpha)+R}{\pm i\coth{\alpha}+R}\,j_{\pm}\no\\
=&\frac{1}{1\pm\frac{\tanh{\zR}}{\tanh{\alpha}}}\left(1\mp \frac{\eta\frac{\tanh{\zR}}{\tanh{\alpha}}(\eta\pm R)}{1\mp\eta R}\right)j_{\pm}\,.
\label{eq:Lax-eta-R}
\end{align}
This expression is precisely the $\eta$-deformed Lax pair\cite{Klimcik2} with the spectral parameter $\laR=\tanh{\zR}/\tanh{\alpha}$ and the deformation parameter $\eta$\,. 
Note that the periodicity of the Lax pair (\ref{eq:Lax-eta-R}) is
\begin{align}
\cL^{\sfR}(\zR+\pi i)=\cL^{\sfR}(\zR)\,.
\end{align}
The associated action can be obtained by setting $\tilde{j}=j$ in (\ref{biaction}) as
\begin{align}
S[g]=&(1+\eta^2)\int_{\cM} d\tau \wedge d\sigma\left\langle j_{-}, \frac{1}{1-\eta R}\,j_{+} \right\rangle\,.
\label{eq:YB-sigma action2}
\end{align}
This expression is again equivalent to the action (\ref{rational action}) for the $\eta$-deformed PCM \cite{Klimcik1,Klimcik2}.

\medskip

\paragraph{The $\mathfrak{su}(2)$ case} As an example, let us consider the case of the Lie algebra $\mathfrak{g}=\mathfrak{su}(2)$\,.
The generators $T^a\,(a=1,2,3)$  of $\mathfrak{su}(2)$ are introduced as
\begin{align}
[T^a,T^b]=\varepsilon^{abc}\, T^c\,, \qquad \Tr(T^{a}T^{b})=-\frac{1}{2}\delta^{ab}\,,\label{su(2) notation}
\end{align}
where $\varepsilon^{abc}$ is a totally antisymmetric tensor normalized as $\varepsilon^{123}=+1$.
The left-invariant 1-form $j$ is then expanded as 
\begin{align}
j_{\pm}=j_{\pm}^{+}T^{-}+j_{\pm}^{-}T^{+}+j_{\pm}^{3}T^{3}\,,
\end{align}
where $T^{\pm}$ are linear combinations of $T^1$ and $T^2$ defined as
\begin{align}
T^{\pm}=\frac{1}{\sqrt{2}}\left(T^1\pm iT^2\right)\,,\qquad
[T^{+},T^{-}]=2iT^{3}\,,\qquad [T^\pm,T^3]=\pm i T^{\pm}\,.\label{su(2) lc-notation}
\end{align}
Let us take the $R$-operator of the Drinfeld-Jimbo type  \cite{Drinfeld,Jimbo} 
such that
\begin{align}
R(T^{\pm})=\mp i T^{\pm},\qquad R(T^3)=0\,.
\label{eq:R-su(2)}
\end{align}
Then, the Lax pair (\ref{eq:Lax-eta-R}) can explicitly be rewritten as
\begin{align}
\mathcal{L}_{\pm}^{\sfR}(\zR)=\frac{\sinh \alpha}{\sinh (\alpha\pm \zR)}
\left[T^{-} e^{-\zR} j_{\pm}^{+}+T^{+} e^{+\zR} j_{\pm}^{-}+\frac{\cosh (\alpha \pm \zR)}{\cosh \alpha}T^{3} j_{\pm}^{3}\right]\,.
\label{eq:rescaled-LaxS3}
\end{align}
The Lax pair (\ref{eq:rescaled-LaxS3}) takes the same expression as (4.22) in \cite{KMY-monod}.

\subsubsection*{ii) Solution with $\tilde{j}=\exp(\pi \Sigma)\, j$}

Next, let us consider the case $\tilde{j}=\exp(\pi \Sigma)\, j$\,.
In this case, the Lax pair (\ref{eq:biLax}) takes the form
\begin{align}
\cL_{\pm}^{\sfR}(\zR)&=
\frac{1}{2\left(1\pm\frac{\tanh{\zR}}{\tanh{\alpha}} \right)}\left(1\mp \frac{\eta\frac{\tanh{\zR}}{\tanh{\alpha}}(\eta\pm R)}{1\mp\eta R}+\frac{1}{\cosh{\zR}}\right)\,j_{\pm}\no\\
&\quad +\frac{1}{2\left(1\pm\frac{\tanh{\zR}}{\tanh{\alpha}} \right)}\left(1\mp \frac{\eta\frac{\tanh{\zR}}{\tanh{\alpha}}(\eta\pm R)}{1\mp\eta R}-\frac{1}{\cosh{\zR}}\right)\,\exp(\pi \Sigma)j_{\pm}\,.
\label{eq:Lax-case2}
\end{align}
Note that the periodicity of the Lax pair (\ref{eq:Lax-case2}) is
\begin{align}
\cL_{\pm}^{\sfR}(\zR+2\pi i)=\cL_{\pm}^{\sfR}(\zR)\,.
\end{align}

\medskip

To rewrite the $2$D action (\ref{biaction}), we expand the left-invariant current $j_{\pm}$ as
\begin{align}
j_{\pm}=\sum_{a}j_{\pm}^{a}H_{a}+\sum_{\alpha>0}\left(j_{\pm}^{\alpha}E_{\alpha}+j_{\pm}^{-\alpha}E_{-\alpha}\right)\,.
\end{align}
By substituting this expansion into (\ref{biaction}) and taking the $R$-operator of 
the Drinfeld-Jimbo type \cite{Drinfeld,Jimbo}
\begin{align}
R(E_{\pm\alpha})=\mp i E_{\pm\alpha}\,, \quad R(H_a)=0\,,
\end{align}
 we obtain
\begin{align}
S[g]&=\int_{\cM}\!d\tau\wedge d \sigma
\biggl(\sum_{\substack{\alpha,\alpha'>0 \\ k_\alpha:\mathrm{odd}}}\left\langle j_{+}^{\alpha}E_{\alpha}\!+\!j_{+}^{-\alpha}E_{-\alpha},j_{-}^{\alpha'}E_{\alpha'}\!+\!j_{-}^{-\alpha'}E_{-\alpha'} \right\rangle + (1+\eta^2) \sum_{a}\left\langle j_{+}^{a}H_{a}, j_{-}^{a}H_{a}     \right\rangle\no\\
&\hspace{85pt}+\sum_{\substack{\alpha,\alpha'>0 \\ k_\alpha:\mathrm{even}}}
\Bigl((1+i\eta)\left\langle j_{+}^{-\alpha}E_{-\alpha},j_{-}^{\alpha'}E_{\alpha'} \right\rangle
+(1-i\eta)\left\langle j_{+}^{\alpha}E_{\alpha},j_{-}^{-\alpha'}E_{-\alpha'} \right\rangle\Bigr)\biggr)\no\\
&=\int_{\cM}d\tau\wedge d \sigma
\biggl(\left\langle j_{+},j_{-} \right\rangle+\eta^2 \sum_{a}\left\langle j_{+}^{a}H_{a}, j_{-}^{a}H_{a} \right\rangle \no\\ 
&\hspace{85pt}-i\eta\sum_{\alpha,\alpha'>0} \left(\left\langle j_{+}^{\alpha}E_{\alpha}+j_{+}^{-\alpha}E_{-\alpha},
 \frac{\exp(\pi\Sigma)+1}{2}(j_{-}^{-\alpha'}E_{-\alpha'}-j_{-}^{\alpha'}E_{\alpha'}) \right\rangle \right) \biggr)\,.
\label{eq:action2-general}
\end{align}
Here we have used the fact that
\begin{align}
\langle E_\alpha,E_\beta \rangle =\delta_{\alpha+\beta,0}\,.
\end{align}
Note that in the second line of the action (\ref{eq:action2-general}),
$
(\exp(\pi\Sigma)+1)/2
$
is a projection to the space where $k_{\alpha}$ is even.

\medskip

It is convenient to rewrite the action (\ref{eq:action2-general}) 
in terms of the $R$-operator. 
To this end, by using the relation
\begin{align}
\left(\frac{1}{1-\eta\,R}+\frac{1}{1+\eta\,R}\right)\, j_{\pm}=\frac{2}{1+\eta^{2}}\sum_{\alpha>0}
\left(j_{\pm}^{\alpha}\,E_{\alpha}+j_{\pm}^{-\alpha}\,E_{-\alpha}\right)+2 \sum_{a}\,j_{\pm}^{a}\,H_{a}\,,
\end{align}
the above action (\ref{eq:action2-general}) can be rewritten as
\begin{align}
S[g]=&\frac{1+\eta^2}{2}\int_{\cM}d\tau\wedge d \sigma\, 
\biggl[\left\langle j_{+},\frac{1}{1+\eta R}j_{-}\right \rangle+\left\langle j_{+},\frac{1}{1-\eta R}j_{-} \right\rangle\biggr]\no\\
&-\eta \int_{\cM}d\tau\wedge d \sigma\, \left\langle j_{+}, \frac{\exp(\pi\Sigma)+1}{2}R\, j_{-}   \right\rangle \,.
\label{eq:2DactionR-double}
\end{align}
The integrability of the system (\ref{eq:2DactionR-double}) is ensured by construction. 
For a direct proof of the integrability, see Appendix \ref{sec:3.80}.
The $B$-field appears only in the second line of the action (\ref{eq:2DactionR-double}) 
since $\Sigma$ and $R$ is skew-symmetric operators by definition. 
It should be remarkable that this action is different from the usual YB-deformation of PCM, 
and this should be a new type of YB deformation. 
Usually, only the factor $1/(1-\eta R)$ is utilized, but here $1/(1+\eta R)$ appears as well. 
The target-space metric obtained from this action is the same as the usual one because 
the metric depends only on $\eta^2$\,. The coupling to the $B$-field is 
also different because it depends on the new ingredient $\Sigma$\,. 

\medskip 

As a remark, it may be interesting to compare the overall factors of (\ref{eq:YB-sigma action2}) 
and (\ref{eq:2DactionR-double})\,. The extra factor 2 is multiplied in  (\ref{eq:YB-sigma action2}) 
in comparison to  (\ref{eq:2DactionR-double})\,. If we consider the solution i) as the deformation 
by two $1/(1-\eta R)$'s, then this factor 2 can be naturally explained. Namely, 
in the solution i), one should have appreciated
\begin{eqnarray}
\frac{2}{1-\eta R} = \frac{1}{1-\eta R} + \frac{1}{1-\eta R}\,, 
\end{eqnarray}
and in the solution ii), one of them is replaced by $1/(1+\eta R)$\,. 
This property would deserve to be called ``chirality''. 

\medskip

\paragraph{The $\mathfrak{su}(2)$ case} 
For completeness, we will give explicit expressions of the Lax pair and the action for the $\mathfrak{su}(2)$ case.
In this case, the operator $\Sigma$ may be identified with the $R$-operator which appeared in the mCYBE (\ref{mCYBE}) since the $\mathfrak{su}(2)$ algebra does not have non-Cartan generators with even $k_{\alpha_j}$.
The Lax pair for $\tilde{j}=\exp(\pi R)\, j$ can be obtained by using (\ref{S3Lax}) as
\begin{align}
\mathcal{L}_{+}^{\sfR}(\wR)&=\frac{1}{e^{2\alpha}\wR^2-1}\left[(e^{2\alpha}-1)(j_{+}^{+}T^{-}+j_{+}^{-}T^{+})\wR+2\tanh{\alpha} \,j_{+}^{3}T^{3}\right]+\tanh{\alpha}\,j_{+}^{3}T^{3}\nonumber\\
&=\frac{(e^{2\alpha}-1)\wR}{e^{2\alpha}\wR^2-1}(j_{+}^{+}T^{-}+j_{+}^{-}T^{+})+\frac{(e^{2\alpha}\wR^2+1)}{(e^{2\alpha}\wR^2-1)}\frac{(e^{2\alpha}-1)}{(e^{2\alpha}+1)}j_{+}^{3}T^{3}\,,\\
\mathcal{L}_{-}^{\sfR}(\wR)&=\frac{1}{\wR^2-e^{2\alpha}}\left[-(e^{2\alpha}-1)(j_{-}^{+}T^{-}+j_{-}^{-}T^{+})\wR-2e^{2\alpha}\tanh{\alpha}\, j_{-}^{3}T^{3}\right]-\tanh{\alpha}\,j_{-}^{3}T^{3}\nonumber\\
&=-\frac{(e^{2\alpha}-1)\wR}{\wR^2-e^{2\alpha}}(j_{-}^{+}T^{-}+j_{-}^{-}T^{+})-\frac{(\wR^2+e^{2\alpha})}{(\wR^2-e^{2\alpha})}\frac{(e^{2\alpha}-1)}{(e^{2\alpha}+1)}j_{-}^{3}T^{3}\,.
\end{align}
They can be expressed in terms of $\zR=\log \wR$ as
\begin{align}
\mathcal{L}_{\pm}^{\sfR}(\zR)=\frac{\sinh \alpha}{\sinh (\alpha\pm \zR)}\left[T^{-} j_{\pm}^{+}+T^{+} j_{\pm}^{-}+\frac{\cosh (\alpha \pm \zR)}{\cosh \alpha}T^{3} j_{\pm}^{3}\right]\,.\label{FR Lax}
\end{align}
The action is also determined as,
\begin{align}
S[g]=-\int_{\cM}d\tau \wedge d\sigma \;\eta^{ij}\left[\mathrm{Tr}(j_i j_j) - 2\eta^2\,\mathrm{Tr}(T^3j_i)\mathrm{Tr}(T^3j_j)\right]\,.\label{sym action}
\end{align}
The Lax pair (\ref{FR Lax}) and the action (\ref{sym action}) are the ones 
for the squashed sigma model \cite{Cherednik,Faddeev:1985qu}.

\medskip

Finally, let us note that the above action can be rewritten into a dipole-like form.
To see this, we introduce the deformed currents as
\begin{align}
    J_{\tau}^{\sfL_{\pm}}&=\frac{1}{2}g\cdot\left(\frac{1+\eta^2}{1\mp \eta\,R}j_{+}+\frac{1+\eta^2}{1\pm \eta\,R}j_{-}\right)\cdot g^{-1}\,,\label{eq:Jt}\\
    J_{\sigma}^{\sfL_{\pm}}&=\frac{1}{2}g\cdot \left(\frac{1+\eta^2}{1\mp \eta\,R}j_{+}-\frac{1+\eta^2}{1\pm \eta\,R}j_{-}\right)\cdot g^{-1}\,.\label{eq:Jx}
\end{align}
By using the action (\ref{eq:R-su(2)}) of the $R$-operator, these deformed currents are 
expressed as\footnote{$ J_{i}^{\sfL_{\pm}}$ correspond to $j^{L_{\pm}}_{\mu}$ 
in \cite{KMY-monod}.}
\begin{align}
     J_{i}^{\sfL_{\pm}}=j_{i}-2\eta^2\Tr(j_{i}T^3)g\cdot T_3\cdot g^{-1}\mp \eta\, \varepsilon_{ij}\partial^{j}(g\cdot T_3\cdot g^{-1})\,,\label{eq:Jpm-2}
\end{align}
where $\varepsilon_{ij}$ is the anti-symmetric tensor and normalized as $\varepsilon_{\tau\sigma}=1$\,. 
By using the expression (\ref{eq:Jpm-2}), we can obtain
\begin{align}
S[g]=-\frac{1}{1+\eta^{2}}\int_{\cM}d\tau\wedge d\sigma \;\eta^{ij}\,\mathrm{Tr}\left(J^{\sfL_{+}}_i J^{\sfL_-}_j\right)\,.\label{dipole-action}
\end{align}

\section{The left-right duality}

In this section, we shall discuss the left-right duality in the $\eta$-deformed PCM.

\medskip

As mentioned previously, the space described by $\wR$ 
is different from the one of $\zL$\,. However, the spectral parameters are related 
through a M\"obius transformation \cite{KMY-monod} 
\begin{align}
\frac{1}{\wR^2}=\frac{\zL+i\eta }{\zL-i\eta}\,.
\label{eq:Mobius}
\end{align}
In fact, the transformation (\ref{eq:Mobius}) maps the twist function (\ref{rational}) 
of the rational description to that of the trigonometric description  (\ref{eq:twist-eta-w}) 
like 
\begin{align}
    \omega
    =\frac{1-\zL^2}{\zL^2+\eta^2}\,d\zL=\frac{4\left(e^{2 \alpha}-\wR^{2}\right)
    \left(e^{2 \alpha} \wR^{2}-1\right)}{\left(e^{4 \alpha}-1\right) 
    \wR\left(\wR^{2}-1\right)^{2}}\,d\wR\,,
\end{align}
where we have used $\eta=i\,\tanh{\alpha}$\,, and set $K=1+\eta^2$ for simplicity.
The transformation (\ref{eq:Mobius}) was originally discovered in \cite{KMY-monod} 
to show the left-right duality in the squashed S$^3$ sigma model. 

\medskip

Since the transformation (\ref{eq:Mobius}) contains the square of $\wR$\,, we have to take care about the parameter region of $\wR$\,. Solving (\ref{eq:Mobius}) in terms of $\wR$\,, we obtain
\begin{align}
\wR=
\begin{cases}
\left(\frac{\zLp+i\eta }{\zLp-i\eta}\right)^{-1/2}\qquad &(\Re \wR>0)\\
-\left(\frac{\zLm+i\eta }{\zLm-i\eta}\right)^{-1/2} \qquad &(\Re \wR<0)
\end{cases}\,.
\label{eq:wR-zLpm}
\end{align}
This map implies that there is a branch cut between $+i\eta$ and $-i\eta$ 
on each Riemann sphere parameterized by $\zLpm$.
Namely, $\mathbb{C}^{\times}$ with $\wR$ (or the cylinder with $\zR$) is regarded 
as the space constructed by joining two $\mathbb{C}P^1$'s with $\zLpm$ via the cut. 
In \cite{KMY-monod}, with this global picture of spectral parameter space, 
the left-right duality has been revealed 
at the level of the affine charge algebras for the $\mathfrak{su}(2)$ case.

\medskip

By taking the Lax pair (\ref{eq:Lax-eta-R-w}) in the trigonometric description, 
the monodromy matrix is given by 
\begin{align}
    T^{\sfR}(\wR):=\mathsf{P}\exp\left(\int_{-\infty}^{\infty} d \sigma\,\cL_{\sigma}^{\sfR}(\sigma;\wR)\right)\,,
    \label{def:mono-R}
\end{align}
where the symbol $\mathsf{P}$ denotes the path-ordering as usual.
The $\tau$ and $\sigma$ components of the Lax pair are given by 
\begin{align}
    \cL_{\tau}^{\sfR} =\frac{1}{2}(\cL_{+}^{\sfR}+\cL_{-}^{\sfR})\,,\qquad 
    \cL_{\sigma}^{\sfR} =\frac{1}{2}(\cL_{+}^{\sfR}-\cL_{-}^{\sfR})\,.\label{Lax-left-two}
\end{align}
Here we suppose the boundary condition that the left-invariant $1$-form $j$ 
vanishes at the spacial infinity.
By expanding $T^{\sfR}(\wR)$ around $\wR=0$ and $\infty$\,, the generators 
of a quantum affine algebra $\widehat{U}_{q}(\mathfrak{g}_R)$ can be obtained 
\cite{KMY-cQA}. One can show the {\it global} equivalence at the level of the monodromy matrix 
(or equivalently conserved charges) 
between the trigonometric and rational descriptions by following  \cite{KMY-monod}. 

\medskip

It is also worth mentioning about the {\it local} equivalence at the level of the Lax pair. 
Namely, the Lax pair (\ref{eq:Lax-eta-R}) is related to the Lax pair for the rational description (\ref{rational Lax}) by the standard gauge transformation:
\begin{align}
\mathcal{L}^{\sfR}_{\pm}=&\frac{\zLp}{\zLp\pm 1}\left(1\mp \frac{\eta (\eta \pm R)}{\zLp(1\mp \eta R)} \right) j_{\pm}\no\\
=&g^{-1}\cdot \mathcal{L}^{\sfL_{+}}_{\pm}(\zLp)\cdot g +g^{-1}\partial_\pm g\,,
\end{align}
where we use the relation between the spectral parameters
\begin{align}
\zLp=\frac{\tanh{\alpha}}{\tanh{\zR}}\,.
\end{align}
Note here that only half of the parameter region of $\zR$ 
\begin{align}
-\infty<\operatorname{Re}\zR<\infty\,,\qquad -\frac{\pi}{2}<\operatorname{Im}\zR<\frac{\pi}{2}
\end{align}
is covered while $\zLp$ spans the whole space of $\mathbb{C}\,$.

\medskip

So far, we have discussed the solution i). For the solution ii), we need to consider more carefully. 
This issue is left as a future problem.

\section{Conclusion and Discussion}

In this paper, we have discussed $\eta$-deformations of PCM from the viewpoint of 
a 4D CS theory. In comparison to \cite{DLMV}, our discussion has started with the trigonometric 
description rather than the rational one. A significant difference is the region of the space of spectral 
parameter and in the trigonometric description, the whole region is covered by construction all at once.  
As a result, the well-known $\eta$-deformed PCM action and its Lax pair have been 
successfully reproduced as a trivial choice $ \tilde{j}=j$\,. In addition, by introducing 
the $\Sigma$ map, another solution has been discovered as a byproduct. 
The resulting action is not the usual form of the YB-deformed PCM because 
the factor $1/(1+\eta R)$ is also contained as well as $1/(1-\eta R)$ in a symmetric way 
and the $B$-field depends on the $\Sigma$ map. Hence this should be 
a new-type of YB-deformation. 

\medskip 

It is significant to generalize this new-type of YB-deformation to the symmetric coset case 
and type IIB string theory on AdS$_5\times$S$^5$ by following \cite{DMV-symm,DMV2-IIB}. 
The coupling to the $B$-field is different from the usual YB deformation, 
and so the other components like R-R fields and dilaton other than the metric would be modified 
due to the appearance of the new ingredient $\Sigma$\,. 
We will report some results in another place \cite{Fukushima2}. 

\medskip 

As another direction, it would be nice to consider a connection between our result and 
the $\lambda$-deformation. 
It is well known that the $\eta$-deformed PCM is related to the $\lambda$-model \cite{lambda1,lambda2} 
via the Poisson Lie T-duality \cite{Vicedo,HT}. So it is interesting to discuss our result from the 
point of view of the $\lambda$-model. 
For recent work on $\lambda$-deformed PCM concerning with 4D CS theory, see \cite{Schmidtt:2019otc,Bassi:2019aaf}.

\medskip 

It may also be interesting to try to generalize our results to the hCYBE case. 
In particular, it seems difficult to generalize the $\Sigma$ map to the hCYBE case.
It may be useful to employ a scaling limit as discussed in Appendix D.

\medskip

We hope that our result would shed light on the relation between 
the global structure of the spectral parameter space and the YB deformation.

\subsection*{Acknowledgments}

It is our pleasure to thank T.~Ishii, T.~Matsumoto, Y.~Sekiguchi and B.~Vicedo for useful discussions.  
The work of J.S. was supported in part by Ministry of Science and Technology (project no. 108-2811-M-002-528), National Taiwan University, and  Osaka City University Advanced Mathematical Institute (MEXT Joint Usage/Research Center on Mathematics and Theoretical Physics).
The works of K.Y.\ was supported by the Supporting Program for Interaction-based Initiative 
Team Studies (SPIRITS) from Kyoto University, and JSPS Grant-in-Aid for Scientific Research (B) 
No.\,18H01214. This work was also supported in part by the JSPS Japan-Russia Research 
Cooperative Program.

\appendix

\section*{Appendix}

\section{How to solve the boundary equations of motion}\label{Appendix drinfeld}

In this appendix, we shall explain how to solve the boundary equations of motion 
(\ref{bc-w1}) and (\ref{bc-w-1})\,.
A significant point observed in \cite{DLMV} is that the boundary equations of motion 
can be solved by regarding $(A_{\alpha},\partial_{\xi_{x}} A_{\alpha})$ 
as an element of a Drinfeld double. Before solving the boundary equations of motion, 
we will give a brief review on the Drinfeld double itself 
(for the details, see, for example, \cite{ReidEdwards:2010vp,Lust:2018jsx,Vysoky-thesis}).

\subsection{Drinfeld double}

A Drinfeld double $\mathfrak{d}$ (of a Lie algebra $\mathfrak{g}_s$) is a Lie algebra equipped 
with a symmetric adjoint-invariant non-degenerate inner product 
$\langle \cdot, \cdot \rangle_{\mathfrak{d}}$\,.
The Drinfeld double $\mathfrak{d}$ is a direct product of two subvector spaces $\mathfrak{g}_s$ and $\tilde{\mathfrak{g}}_s$ as a vector space 
\begin{align}
\mathfrak{d}=\mathfrak{g}_s\oplus \tilde{\mathfrak{g}}_s\,, 
\end{align}
where $\mathfrak{g}_s$ and $\tilde{\mathfrak{g}}_s$ are Lie subalgebras of $\mathfrak{d}$ 
with the same dimension $d={\rm dim}\,\mathfrak{g}_s={\rm dim}\,\tilde{\mathfrak{g}}_s$\,.
Let $\{T_a\}$ and $\{\widetilde{T}^a\}\,(a=1,\dots , d)$ be the generators of $\mathfrak{g}_s$ 
and $\tilde{\mathfrak{g}}_s$\,, respectively.
These generators satisfy
\begin{align}
\langle T_a,T_a\rangle_\mathfrak{d} =0\,,\qquad
\langle \widetilde{T}^a,\widetilde{T}^b\rangle_\mathfrak{d} =0\,,\qquad
\langle T_a,T^b\rangle_\mathfrak{d} =\delta_a^b\,. 
\label{eq:inner-double-re}
\end{align}
Namely, $\mathfrak{g}_s$ and $\tilde{\mathfrak{g}}_s$ are the maximal isotropic subalgebras of $\mathfrak{d}$ with respect to the inner product $\langle \cdot, \cdot \rangle_{\mathfrak{d}}$\,.
By defining $T_{A}:=(T_{a}, \widetilde{T}{}^a)$\,,
the relations in (\ref{eq:inner-double-re}) can be recast into a simple form,  
\begin{align}
\langle T_A,T_B\rangle_\mathfrak{d} =\eta_{AB}=
\begin{pmatrix}
~0_{d}~~&~\delta_a{}^b~\\
~\delta^a{}_b~&~0_{d}~
\end{pmatrix}
\,.\label{O(d,d)-metric}
\end{align}
Here $0_d$ denotes the $d \times d$ zero matrix.
This expression indicates that the structure group on the Drinfeld double $\mathfrak{d}$ 
is $O(d,d)$\,.

\medskip

Suppose that the defining relations of $\mathfrak{d}$ are given by
\begin{align}
[T_{A},T_{B}]=F_{AB}{}^{C}T_{C}\,,
\end{align}
where $F_{AB}{}^{C}$ are the structure constants of $\mathfrak{d}$\,.
In terms of $T_{a}$ and $\tilde{T}^{a}$\,, the commutation relations are rewritten as
\begin{align}
[T_a, T_b]=f_{ab}{}^cT_c\,,\qquad
[T_a, \widetilde{T}^b]=\tilde{f}^{bc}{}_{a} T_c-f_{ac}{}^b\widetilde{T}^c\,,\qquad
[\widetilde{T}^a, \widetilde{T}^b]=\tilde{f}^{ab}{}_c\widetilde{T}^c\,,
\label{eq:com-double}
\end{align}
where $f_{ab}{}^c:= F_{ab}{}^{c}$ and $\tilde{f}^{ab}{}_{c}:= F^{ab}{}_{c}$ are 
the structure constants of $\mathfrak{g}_s$ and $\tilde{\mathfrak{g}}_s$\,, respectively.
Furthermore, the Jacobi identity for $\mathfrak{d}$ leads to the following relations 
between $f_{ab}{}^c$ and $\tilde{f}^{ab}{}_{c}$~:
\begin{align}
\tilde{f}^{ce}{}_{d}f_{ab}{}^{d}=4\,\tilde{f}^{d[c}{}_{[a}f_{b]d}{}^{e]}\,.\label{eq:Jacobi-double}
\end{align}

\medskip

By definition, the Drinfeld double $\mathfrak{d}$ has a decomposition into two Lie subalgebras 
$\mathfrak{g}$ and $\tilde{\mathfrak{g}}$ satisfying (\ref{O(d,d)-metric}), (\ref{eq:com-double}) and (\ref{eq:Jacobi-double}).
The triple pair $(\mathfrak{d}, \mathfrak{g}_s, \tilde{\mathfrak{g}}_s)$ is called a Manin triple.
In general, a given Drinfeld double $\mathfrak{d}$ can have some Manin triples, namely, 
\begin{align}
    \mathfrak{d}=\mathfrak{g}_{s}\oplus \tilde{\mathfrak{g}}_{s}=\mathfrak{g}_{s}'\oplus \tilde{\mathfrak{g}}_{s}'=\cdots\,,
\end{align}
where each of the Manin triples satisfies the conditions (\ref{O(d,d)-metric}), (\ref{eq:com-double}) and (\ref{eq:Jacobi-double})\,.

\subsection{Solutions to the boundary equations of motion (\ref{bc-w1}), (\ref{bc-w-1}) }

Let us solve the boundary equations of motion 
\begin{align}
   \epsilon^{ij} \llangle(A_{i},\partial_{\xi_{p}} A_{i}),\delta(A_{j}, \partial_{\xi_{p}} A_{j})\rrangle_{\mathfrak{t},p}=0\,, \qquad p\in\mathfrak{p}\,,
\end{align}
where the double bracket is defined as
\begin{align}
    \llangle (x,y), (x',y')\rrangle_{\mathfrak{t},p}&:=(\text{res}_{x}\omega)\langle x, x'\rangle+(\text{res}_{p}\,\xi_{p}\omega)\left(\langle x, y'\rangle+\langle x', y\right\rangle)\no\\
    &=K\left(\langle x, y'\rangle+\langle x', y\rangle\right)\,.
    \label{def:inner-double}
\end{align}
As discussed in \cite{Klimcik:2018vhl,DLMV},  $(A_{i},\partial_{\xi_{p}} A_{i})$ can be regarded as an element of a Drinfeld double with the inner product $\llangle \cdot, \cdot\rrangle_{\mathfrak{t},p}$, and hence we can solve the boundary equation of motion. 
In the following discussion, we will explain this point.

\subsubsection*{i) Semi-abelian double}

By definition, the double bracket is a symmetric non-degenerate inner product on a vector space $\mathfrak{t}$ which is isomorphic to the direct product of two $\mathfrak{g}_s$\,,
\begin{align}
    \mathfrak{t}&=\mathfrak{k}\oplus\tilde{\mathfrak{k}}\,,\\
    \mathfrak{k}&=\{(x,0)|x\in\mathfrak{g}_s \},\\
    \tilde{\mathfrak{k}}&=\{(0,y)|y\in\mathfrak{g}_s \}\,.
\end{align}
Let us first discuss the Lie algebraic structure of $\mathfrak{t}$ with a group multiplication of $G_s\ltimes \mathfrak{g}_s$ which has $(g,\mathcal{A}):= (g,g\partial_{\xi}g^{-1})$ as an element.
The multiplication rule of $G_s\ltimes \mathfrak{g}_s$ is induced by a group multiplication of $G_s$\,.

\medskip

To this end, let us consider 
\begin{equation}
G_s\times G_s \to G_s: \quad (g_1, g_2) \mapsto g_1\cdot  g_2\,. 
\end{equation} 
Then, we obtain
\begin{align}
    (g_1, \mathcal{A}_1)\cdot  (g_2, \mathcal{A}_2)=(g_1\cdot g_2, \mathcal{A}_1+{\rm Ad}_{g_1}(\mathcal{A}_2) )\,,
\end{align}
where $\mathcal{A}_i=\partial_{\xi} g_ig^{-1}_i\,(i=1,2)$ and ${\rm Ad_{g}}(x)= g\cdot  x\cdot  g^{-1}$ for $x\in\mathfrak{g}_s$\,.
By using this rule, the inverse of $(g, \mathcal{A})$ is given by
\begin{align}
    (g, \mathcal{A})^{-1}=(g^{-1}, -{\rm Ad}_{g^{-1}}(\mathcal{A}))\,.
\end{align}
Then, the right-invariant current is
\begin{align}
    -d(g,\mathcal{A})\cdot (g, \mathcal{A})^{-1}=-(dgg^{-1},d\mathcal{A}+[\mathcal{A},dg g^{-1}])=(A, \partial_{\xi}A)\,,
\end{align}
and the adjoint action is given by
\begin{align}
    {\rm Ad_{(h,\partial_{\xi}hh^{-1})}}\left((A, \partial_{\xi}A)\right)
    =({\rm Ad_{h}}(A), {\rm Ad}_{h}(\partial_{\xi}A)+[\partial_{\xi}hh^{-1},{\rm Ad}_{h}A])\,,
\end{align}
where $h\in G_s$\,.
This adjoint action implies that the vector space $\mathfrak{t}$ has the following Lie algebra commutator
\begin{align}
    [(x,y), (x',y')]_{\mathfrak{t}}= ([x,x'], [x,y']-[x',y])\,,
    \label{eq:semi-ab-com}
\end{align}
and the inner product (\ref{def:inner-double}) is adjoint invariant.
In fact, 
\begin{align}
    &\llangle  {\rm Ad}_{(h,\partial_{\xi}hh^{-1})}((x_1,y_1)), {\rm Ad}_{(h,\partial_{\xi}hh^{-1})}((x_2, y_2))\rrangle_{\mathfrak{t},p}\no\\
    &=K(\langle {\rm Ad}_{h}x_1,{\rm Ad}_{h}y_2+[\partial_{\xi}hh^{-1},{\rm Ad}_{h}x_2 ]   \rangle+\langle {\rm Ad}_{h}y_1+[\partial_{\xi}hh^{-1},{\rm Ad}_{h}x_1 ],{\rm Ad}_{h}x_2 \rangle)\no\\
     &=K(\langle x_1,y_2 \rangle+\langle y_1,x_2 \rangle+ \langle x_1, [h^{-1}\partial_{\xi}h,x_2 ]  \rangle+ \langle  [h^{-1}\partial_{\xi}h,x_1 ],x_2  \rangle  )\no\\
    &=  \llangle  (x_1,y_1), (x_2, y_2)\rrangle_{\mathfrak{t},p}\,.
\end{align}
In the final equation, we have used the fact that the inner product $\langle \cdot,\cdot \rangle$ is adjoint invariant.

\medskip

We will consider the inner product (\ref{def:inner-double}) and the commutator (\ref{eq:semi-ab-com}) in more detail.
It is convenient to introduce the generators $T_{a}$ and $\tilde{T}^{a}$ of two vector subspaces $\mathfrak{k}$ and $\tilde{\mathfrak{k}}$\,, respectively. They are represented by the generators $t_{a}\,(a=1,\dots,d:=\text{dim}\,\mathfrak{g}_s)$ of $\mathfrak{g}_s$ as
\begin{align}
    T_{a}=(t_{a},0)\,,\qquad \tilde{T}^{a}=(0,t^{a})\,.
\end{align}
Here $t^{a}:= t_{b}\eta^{ab}$\,, where $\eta_{ab}$ is the Killing form of $\mathfrak{g}_s$\,, 
and $t_{a}$'s are normalized as
\begin{align}
    \left\langle t_{a}, t_{b}\right\rangle=\eta_{ab}\,.\label{tt-inner}
\end{align}
Then, the generators $T_{A}=(T_{a},\tilde{T}^{a})$ satisfy 
\begin{align}
   \llangle T_{A}, T_{B}\rrangle_{\mathfrak{t},x}=K\eta_{AB}\,,
    \label{eq:double-metric-CS}
\end{align}
where $\eta_{AB}$ is defined in (\ref{O(d,d)-metric}).
Equivalently, $\mathfrak{k}$ and $\tilde{\mathfrak{k}}$ are maximally isotropic with respect to the inner product $\llangle \cdot\,,\cdot \rrangle_{\mathfrak{t},x}$.
The commutation relations of $T_{A}=(T_{a},\tilde{T}^{a})$ are given by
\begin{align}
    [T_{a},T_{b} ]_{\mathfrak{t}}= f_{ab}{}^c\,T_{c}\,,\qquad
    [\tilde{T}^{a},\tilde{T}^{b}]_{\mathfrak{t}}=0\,,\qquad
    [T_{a},\tilde{T}^{b}]_{\mathfrak{t}}=-f_{ac}{}^{b}\tilde{T}^{c}\,,
\end{align}
where $f_{ab}{}^c$ are the structure constants of $\mathfrak{g}_s$\,.
This implies
\begin{align}
    \mathfrak{k}=\mathfrak{g}_s\ltimes \{0\}\,,\qquad 
    \tilde{\mathfrak{k}}=\{0\}\ltimes \mathfrak{g}_{\text{ab}}\,,
    \label{eq:abelian-double}
\end{align}
where $\mathfrak{g}_{s,\text{ab}}$ is an abelian algebra with $\text{dim}\,\mathfrak{g}_{s,\text{ab}}=d$\,.
Furthermore, the generators $T_{A}$'s satisfy the Jacobi identity
\begin{align}
    [[T_{A},{T}_{B}],T_{C}]+[[T_{B},{T}_{C}],T_{A}]+[[T_{C},{T}_{A}],T_{B}]=0\,.
\end{align}
Therefore, $\mathfrak{t}$ is a Drinfeld double with a Manin triple $(\mathfrak{t},\mathfrak{k},\tilde{\mathfrak{k}})$ and the inner product (\ref{def:inner-double}). The Drinfeld double $\mathfrak{t}$ is often called a semi-abelian double.

\medskip

As a result, we can solve the boundary equation of motion by requiring
\begin{align}
    (A_{i},\partial_{\xi_{p}} A_{i})\in \mathfrak{k}\qquad \text{or}\qquad (A_{i},\partial_{\xi_{p}} A_{i})\in \tilde{\mathfrak{k}}\,.
\end{align}
If we take the second boundary condition, we obtain the PCM with $G_s$.   

\subsubsection*{ii) Other solutions}

The Drinfeld double $\mathfrak{t}$ can have other Manin triples.
An important thing is the Manin triple $(\mathfrak{t},\mathfrak{g}_s\ltimes \{0\}, \mathfrak{g}_{s,R})$ with the commutation relations
\begin{align}
\begin{split}
    &[T'_{a},T'_{b} ]_{\mathfrak{t}}= f_{ab}{}^c\,T'_{c}\,,\qquad
    [\tilde{T}'{}^{a},\tilde{T}'{}^{b}]_{\mathfrak{t}}=\tilde{f}^{ab}{}_{c}\tilde{T}'{}^{c}\,,\\
    &[T'_{a},\tilde{T}'{}^{b}]_{\mathfrak{t}}=\tilde{f}^{cb}{}_{a}T'_{c}-f_{ac}{}^{b}\tilde{T}'{}^{c}\,,
\end{split}
\end{align}
where $T'_{a}$ and $\tilde{T}'{}^{a}$ are the generators of $\mathfrak{g}_s\ltimes \{0\}$ and 
$ \mathfrak{g}_{s,R}$\,, respectively, and $\tilde{f}^{ab}{}_{c}$ are the structure constants of $\mathfrak{g}_{s,R}$ defined as
\begin{align}
\tilde{f}^{ab}{}_{c}=\eta\,r^{ad}f_{dc}{}^{b}-\eta\,r^{bd}f_{dc}{}^{a}\,.
\end{align}
Here $\eta$ is a real parameter. When $\eta=0$\,, the above Manin triple reduces to the previous one (\ref{eq:abelian-double}).
The skew-symmetric constant matrix $r^{ab}=-r^{ba}$ satisfies the hCYBE
\begin{align}
 f_{e_1e_2}{}^a\,r^{be_1}\,r^{ve_2} + f_{e_1e_2}{}^b\,r^{ce_1}\,r^{ae_2} + f_{e_1e_2}{}^c\,r^{ae_1}\,r^{be_2} =0\,,
\label{eq:CYBE-r}
\end{align}
and gives rise to a classical $r$-matrix $r\in\mathfrak{g}_s\otimes \mathfrak{g}_s$ 
in the tensorial notation
\begin{align}
    r=\frac{1}{2}r^{ab}t_{a}\wedge t_{b}=\frac{1}{2}r^{ab}(t_{a}\otimes t_{b}-t_{b}\otimes t_{a})\,.
    \label{eq:hCYBE-tensor}
\end{align}
The hCYBE ensures that $\tilde{f}^{ab}{}_{c}$ satisfies the Jacobi identity.

\medskip

Remarkably, the two Manin triples $(\mathfrak{t},\mathfrak{k},\tilde{\mathfrak{k}})$ and $ (\mathfrak{t},\mathfrak{g}_s\ltimes \{0\}, \mathfrak{g}_{s,R})$ are related by an $O(d,d)$ transformation\footnote{This relation has been observed in the classification of six-dimensional Drinfled doubles\cite{Hlavaty:2002kp,Snobl:2002kq}},
\begin{align}
\begin{split}
    T'_{A}&=T_{B}\,\mathcal{O}^{B}{}_{A}\,,\\
    \mathcal{O}^{B}{}_{A}&=\begin{pmatrix}
    \mathcal{O}^{b}{}_{a}&\mathcal{O}^{ba}\\
    \mathcal{O}_{ba}&\mathcal{O}_{b}{}^{a}
    \end{pmatrix}
    =\begin{pmatrix}
    \delta^{b}_{a}&\eta\,r^{ba}\\
    0_{d}&\delta_{b}^{a}
    \end{pmatrix}\in \mathcal{O}(d,d)\,,
    \end{split}
    \label{eq:O(d,d)-tr}
\end{align}
or equivalently,
\begin{align}
    T'_{a}=T_{a}\,,\qquad \tilde{T}'{}^{a}=\tilde{T}^{a}+\eta\,T_{b}r^{ba}\,.
    \label{eq:O(d,d)-beta-T}
\end{align}
The transformation (\ref{eq:O(d,d)-tr}) is a $\beta$-transformation acting on the generators of $\mathfrak{t}$\,, and preserves the $O(d,d)$ metric (\ref{eq:double-metric-CS}).
This observation leads to explicit elements of the Lie algebra $\mathfrak{g}_{s,R}$\,.
By using the transformation rule (\ref{eq:O(d,d)-beta-T}), the deformed dual generator $\tilde{T}'{}^{a}$ is
\begin{align}
    \tilde{T}'{}^{a}=\tilde{T}^{a}+\eta\,T_{b}r^{ba}=(0,t^{a})+(\eta R(t^{a}),0)=(\eta R(t^{a}),t^{a})\,,
\end{align}
where the $R$-operator $R:\mathfrak{g}_s\to \mathfrak{g}_s$ is defined as
\begin{align}
    R(x):= \frac{1}{2} r^{ab}(t_{a}\langle t_{b},x \rangle-t_{b}\langle t_{a},x \rangle)\,,\qquad x\in\mathfrak{g}_s.
\end{align}
In terms of the $R$-operator, the hCYBE (\ref{eq:hCYBE-tensor}) can be rewritten as
\begin{align}
   \text{CYBE}(x,y)= [R(x),R(y)]-R([R(x),y]+[x,R(y)])=0\,,\qquad x\,,y\in\mathfrak{g}_s\,.
    \label{eq:hCYBE}
\end{align}
As a result, the dual Lie algebra $\mathfrak{g}_{s,R}$ is represented by
\begin{align}
    \mathfrak{g}_{s,R}=\{(\eta R(x),x)\,|\,x\in\mathfrak{g}_s\}\,.
    \label{def;gR}
\end{align}
In this way, the Manin triple $(\mathfrak{t},\mathfrak{g}_s\ltimes \{0\}, \mathfrak{g}_{s,R})$ is generated from $(\mathfrak{t},\mathfrak{k},\tilde{\mathfrak{k}})$ by the $O(d,d)$ transformation (\ref{eq:O(d,d)-tr}).
In particular, this fact means that the homogeneous YB deformations can be regarded as $\beta$-transformations \cite{Sakamoto:2017cpu,Sakamoto:2018krs,Lust:2018jsx}.

\medskip

As in the previous case, the boundary equations of motion can be solved by taking the boundary conditions,
\begin{align}
    (A_{i},\partial_{\xi_{p}} A_{i})\in (\mathfrak{g}_s\ltimes \{0\})\qquad \text{or}\qquad (A_{i},\partial_{\xi_{p}} A_{i})\in \mathfrak{g}_{s,R}\,.
\end{align}
The second choice gives rise to the homogeneous YB deformation of the $G_s$-PCM associated with the classical $r$-matrix (\ref{eq:CYBE-r}).

\subsection{A solution to the boundary equation of motion (\ref{eq:bEOM-mCYBE})}

Finally, we will consider the direct product $\mathfrak{h}_s$ composed of vector spaces 
$\mathfrak{g}_s^\delta$ and $\mathfrak{g}_{s,R}$ like 
\begin{align}
    \mathfrak{h}_s&:=\mathfrak{g}_s^{\delta}\oplus \mathfrak{g}_{s,R}\,,\\
\mathfrak{g}_s^\delta&:=\left\{(x,x)|x\in\mathfrak{g}\right\}\,,\\
\mathfrak{g}_{s,R}&:= \left\{((R-i)x,(R+i)x)|x\in\mathfrak{g}\right\}\,.
\end{align}
The linear $R$-operator is associated with the classical $r$-matrix of Drinefeld-Jimbo 
type for $\mathfrak{g}$ and satisfies the modified CYBE (\ref{mCYBE}) with $c=i$\,. 
The vector spaces $\mathfrak{h}_s$\,, $\mathfrak{g}_s^\delta$ and $\mathfrak{g}_{s,R}$
are real forms of (\ref{def:d=gd+gR}), (\ref{def:gR}) and (\ref{def:gd}). 
Then (\ref{def:gR}) can be used as a solution to the boundary e.o.m. (\ref{eq:bEOM-mCYBE}).
In the following discussion, we will show that $\mathfrak{h}_s$ can be seen as a Drinfeld double.

\medskip

To begin with, let us take the following basis elements for two vector subspaces 
$\mathfrak{g}_s^{\delta}\,, \mathfrak{g}_{s,R}$ as, respectively, 
\begin{align}
    T_a=(t_a,t_a)\,,\qquad  \widetilde{T}^a=((R-i)t^a,(R+i)t^a)\,. 
    \label{d-TtT}
\end{align}
Here $t^a$ are generators of $\mathfrak{g}$ satisfying (\ref{tt-inner}).
The inner product on $\mathfrak{h}_s$ is defined as
\begin{align}
 \llangle (x,y),(x',y')  \rrangle_{\mathfrak{h}_s}&:=
\frac{i}{2}
(\langle x,x' \rangle - \langle y,y' \rangle ) \,,
\end{align}
which has the same form with (\ref{eq:d-inner}) up to an overall factor.
It is then easy to show that these generators satisfy 
\begin{align}
   \llangle T_{A}, T_{B}\rrangle_{\mathfrak{h}_s}=\eta_{AB}\,.
\end{align}
Therefore, $\mathfrak{g}_s^{\delta}$ and $\mathfrak{g}_{s,R}$ are maximal isotropic subalgebras of $\mathfrak{h}_s$ with respect to the inner product $\llangle \cdot, \cdot \rrangle_{\mathfrak{h}_s}$\,.

\medskip

Next, let us show that $\mathfrak{h}_s$ has a Lie-algebraic structure, or equivalently $\mathfrak{g}_s^{\delta}$ and $\mathfrak{g}_{s,R}$ are Lie algebras and satisfy the relations (\ref{eq:com-double}) and (\ref{eq:Jacobi-double})\,.
By definition, $T_{a}$'s satisfy the standard relations
\begin{align}
    [T_{a}, T_{b}]_{\mathfrak{h}_s}=f_{ab}{}^{c}T_{c}\,,
\end{align}
where the commutator $ [\cdot, \cdot ]_{\mathfrak{h}_s}$ is defined as
\begin{align}
 [(x,y), (x',y') ]_{\mathfrak{h}_s}:=([x,x'],[y,y'])\,.
\end{align}
The commutator for $\widetilde{T}^a$'s is given by 
\begin{align}
    [\widetilde{T}^a, \widetilde{T}^b ]_{\mathfrak{h}_s}&=\left([(R-i)t^a, (R-i)t^b], [(R+i)t^a, (R+i)t^b]\right)\,.\label{com:tTtT}
\end{align}
Each entry in (\ref{com:tTtT}) is evaluated as
\begin{align}
    [(R\pm i)t^a, (R\pm i)t^b]&=[R(t^a),R(t^b)]\pm i [t^a,(R\pm i)t^b]\pm i [t^a,R(t^b)]-[t^a,t^b]\no\\
    &=(R\pm i)\left([R(t^a),t^b]+ [t^a,R(t^b)]  \right)\no\\
    &=(r^{ac}f_{cd}{}^{b}-r^{bc}f_{cd}{}^{a})(R\pm i)(t^d)\,,
\end{align}
where in the second equality, the mCYBE (\ref{mCYBE}) has been used.
The matrix components of the $R$-operator, $r^{ab}$ are given by the relation 
\begin{equation}
R(t^a)=-r^{a}{}_{c}\,t^c\,,
\end{equation} 
and satisfy the mCYBE
\begin{align}
 f_{e_1e_2}{}^a\,r^{be_1}\,r^{ce_2} + f_{e_1e_2}{}^b\,r^{ce_1}\,r^{ae_2} + f_{e_1e_2}{}^c\,r^{ae_1}\,r^{be_2} =f^{abc}\,.
\label{eq:mCYBE-r}
\end{align}
Therefore, $\widetilde{T}^a$'s satisfy  
\begin{align}
     [\widetilde{T}^a, \widetilde{T}^b ]_{\mathfrak{h}_s}=\tilde{f}^{ab}{}_{c}\widetilde{T}^c\,,
\end{align}
where the structure constants $\tilde{f}^{ab}{}_{c}$ are given by 
\begin{align}
    \tilde{f}^{ab}{}_{c}=r^{ad}f_{dc}{}^{b}-r^{bd}f_{dc}{}^{a}\,.
    \label{dual:mCYBE}
\end{align}
In particular, thanks to the mCYBE (\ref{eq:mCYBE-r}), the structure constants  
(\ref{dual:mCYBE}) satisfy the Jacobi identity.
Furthermore, by using (\ref{dual:mCYBE})\,, we can obtain
\begin{align}
    [t_a,(R\pm i)t^b]=\tilde{f}^{bc}{}_{a}t_c-f_{ac}{}^{b}(R\pm i)t^c\,.
\end{align}
This relation indicates
\begin{align}
    [T_a ,\widetilde{T}^b  ]_{\mathfrak{h}_s}&= ([t_a,(R-i)t^b], [t_a,(R+i)t^b])=\tilde{f}^{bc}{}_{a}T_{c}-f_{ac}{}^{b}\widetilde{T}{}^{c}\,.
\end{align}
Finally, by using the Jacobi identity of $f_{ab}{}^{c}$\,, we can check that the structure constants $f_{ab}{}^{c}$ and $\tilde{f}^{ab}{}_{c}$ satisfy the relation (\ref{eq:Jacobi-double}).
As a result, $\mathfrak{h}_s$ can be regarded as a Drinfeld double.

\section{Representation of $\lambda$-map}\label{appendix:exp}

Let discuss here a possible construction of a $\lambda$-map in (\ref{eq:generator-map-expR})\,. 

\medskip 

For example, by the use of a Cartan generator (say $H_1$)\,, 
it is possible to construct a semigroup homomorphism $\alpha\mapsto\lambda_{\alpha}$ 
labeled by the integer $k_{\alpha}$ as follows: 
\begin{align}
\lambda_{\alpha}=-i\frac{\alpha(H_1)}{\bar{\alpha}}\pi :=-ik_{\alpha}\pi 
\,,
\end{align}
where $\bar{\alpha}$ is a constant to be fixed appropriately. 
Then the map $\Sigma:\mathfrak{g}^{\mathbb{C}}\rightarrow\mathfrak{g}^{\mathbb{C}}$ is given by
\begin{align}
\Sigma(E_{\alpha}) :=-i \,\frac{\alpha(H_1)}{\bar{\alpha}}\,E_{\alpha}=-i k_{\alpha}E_{\alpha}\,,\qquad 
\Sigma(H_a)=0\,.
\end{align}
One can see that the exponential map $\exp(\pi\Sigma)$ indeed exhibits $\mathbb{Z}_2$-grading.

\medskip

As a concrete example, it is instructive to consider the $\mathfrak{su}(n)$ case. 
Explicitly, the weights for the defining representation of $\mathfrak{su}(n)$ are given 
in the standard form: (for example, see \cite{Georgi})
\begin{align}
\nu^1=&\left(\frac{1}{2},\frac{1}{2\sqrt{3}},\dots,\frac{1}{\sqrt{2m(m+1)}},\dots,\frac{1}{\sqrt{2(n-1)n}} \right)\,,\no\\
\nu^2=&\left(-\frac{1}{2},\frac{1}{2\sqrt{3}},\dots,\frac{1}{\sqrt{2m(m+1)}},\dots,\frac{1}{\sqrt{2(n-1)n}} \right)\,,\no\\
\nu^3=&\left(0,-\frac{1}{\sqrt{3}},\dots,\frac{1}{\sqrt{2m(m+1)}},\dots,\frac{1}{\sqrt{2(n-1)n}} \right)\,,\no\\
\vdots\;&\no\\
\nu^{m+1}=&\left(0,0,\dots,-\frac{m}{\sqrt{2m(m+1)}},\dots,\frac{1}{\sqrt{2(n-1)n}} \right)\,,\no\\
\vdots\;&\no\\
\nu^n=&\left(0,0,\dots,0,\dots,-\frac{n-1}{\sqrt{2(n-1)n}} \right)\,.
\end{align}
Then the simple roots can be obtained from the weights as\footnote{The convention of the positive root in \cite{Georgi} is backward. See Chapter 13 of \cite{Georgi}.} 
\begin{align}
\alpha^{(m)}=\nu^m-\nu^{m+1} \qquad (m = 1,2,\dots,n-1)\,. 
\end{align}
Hence, by taking that $\bar{\alpha}=1/2$\,, 
the first component of arbitrary roots are expressed as 
\[
\alpha(H_1) = m_1 \times 1 + m_2 \times (-1/2) := k_{\alpha}/2 
~(m_1,m_2,k_{\alpha} \in \mathbb{Z})\,.
\]
Thus the $\lambda$-map has been well constructed for the $\mathfrak{su}(n)$ case.

\section{A direct proof of the integrability for (\ref{eq:2DactionR-double})}
\label{sec:3.80}

Here, we shall directly prove the classical integrability of the new system 
(\ref{eq:2DactionR-double}) by showing that the equation of motion of the action 
(\ref{eq:2DactionR-double}) is equivalent to the flatness condition of the Lax pair  
(\ref{eq:Lax-case2}). Although the classical integrability is ensured by construction, 
it is still instructive to see the structure of the Lax pair in detail. 

\subsection{The equation of motion for (\ref{eq:2DactionR-double})}

First of all, let us derive the equation of motion from the classical action 
(\ref{eq:2DactionR-double}). 
Taking an infinitesimal variation
\begin{align}
g\mapsto g +\delta g=g +g\epsilon \quad &\Longrightarrow \quad \delta j= d\epsilon +[ j, \epsilon]\no\\
&\Longleftrightarrow\quad\left\{
\begin{array}{l}
\delta j_+=\pa_+ \epsilon +[j_+,\epsilon]\\
\delta j_-=\pa_- \epsilon +[j_-,\epsilon]
\end{array}\right.
\end{align}
for the action (\ref{eq:2DactionR-double}), we can obtain
\begin{align}\begin{split}
\delta S[g]=& \frac{1+\eta^2}{2}\int_{\cM}d\tau\wedge d\sigma\,
\Bigg[-\Bigg\langle \epsilon,\pa_+\left(\frac{2}{1-\eta^2 R^2}j_{-}\right) +\pa_-\left(\frac{2}{1-\eta^2 R^2}j_{+}\right)\\
&\hspace{140pt}+\left[j_{-},\frac{2}{1-\eta^2 R^2}j_{+}\right] + \left[j_{+}, \frac{2}{1-\eta^2 R^2}j_{-}\right] \Bigg\rangle \Bigg] \\
&-\eta\!\int_{\cM} \!d\tau \wedge d\sigma\big\langle \epsilon, 
-\pa_+ \!\left(\Se Rj_- \right)+ \pa_- \!\left(\Se Rj_+\right)
-\left[ j_+,\Se Rj_-\right] + \left[ j_-,\Se Rj_+\right]
\big\rangle\,.
\end{split}\end{align}
Here we have introduced the following projection operators into the spaces 
with even and odd values of $k_{\alpha}$ in (\ref{integer}), respectively, 
\begin{align}
\Se:=\frac{1+\exp[\pi\Sigma]}{2}\,, \qquad \So:= \frac{1-\exp[\pi\Sigma]}{2}\,.
\end{align}
Thus the equation of motion reads
\begin{align}\begin{split}
0=&(1+\eta^2)\bigg[\pa_+\left(\frac{1}{1-\eta^2 R^2}j_{-}\right) +\pa_-\left(\frac{1}{1-\eta^2 R^2}j_{+}\right)\\
&\hspace{50pt}+\left[j_{-},\frac{1}{1-\eta^2 R^2}j_{+}\right] + \left[j_{+}, \frac{1}{1-\eta^2 R^2}j_{-}\right]\bigg] \\
&-\eta\left[
\pa_+ \left(\Se Rj_- \right)- \pa_- \left(\Se Rj_+\right)
+\left[ j_+,\Se Rj_-\right] - \left[ j_-,\Se Rj_+\right] \right]\,,
\end{split}\end{align}
and it can be decomposed into the even and odd parts:
\begin{align}\begin{split}
\cE=&\;\Ee+\Eo=0\,,
\end{split}\label{eom-even-odd}\end{align}
\begin{align}
\begin{split}
\Ee:=&(1+\eta^2)\bigg[\pa_+\left(\frac{1}{1-\eta^2 R^2}\Se j_{-}\right) +\pa_-\left(\frac{1}{1-\eta^2 R^2}\Se j_{+}\right)\\
&\hspace{48pt}+\left[\Se j_{-},\frac{1}{1-\eta^2 R^2}\Se j_{+}\right] + \left[\Se j_{+}, \frac{1}{1-\eta^2 R^2}\Se j_{-}\right]
\bigg]\\
&-\eta\left[
\pa_+ \left(\Se Rj_- \right)- \pa_- \left(\Se Rj_+\right)
+\left[ \Se j_+,\Se Rj_-\right] - \left[ \Se j_-,\Se Rj_+\right] \right]\,, 
\end{split} 
\label{EOM-even} \\
\begin{split}
\Eo:=&(1+\eta^2)\bigg[\pa_+\left(\frac{1}{1-\eta^2 R^2}\So j_{-}\right) +\pa_-\left(\frac{1}{1-\eta^2 R^2}\So j_{+}\right)\\
&\hspace{47pt}+\left[\So j_{-},\frac{1}{1-\eta^2 R^2}\Se j_{+}\right] + \left[\So j_{+}, \frac{1}{1-\eta^2 R^2}\Se j_{-}\right]
\\
&\hspace{47pt}+\left[\Se j_{-},\frac{1}{1-\eta^2R^2}\So j_{+}\right] + \left[\Se j_{+}, \frac{1}{1-\eta^2R^2}\So j_{-}\right]
\bigg]\\
&-\eta\left[
\left[ \So j_+,\Se Rj_-\right] - \left[ \So j_-,\Se Rj_+\right] \right]\,.
\end{split}\label{EOM-odd}\end{align}
Note here that
\begin{align}\begin{split}
&\left[\So j_{-},\frac{1}{1-\eta^2 R^2}\So j_{+}\right] + \left[\So j_{+}, \frac{1}{1-\eta^2 R^2}\So j_{-}\right]\\
=&\left[\So j_{-},\frac{1}{1+\eta^2}\So j_{+}\right] + \left[\So j_{+}, \frac{1}{1+\eta^2}\So j_{-}\right]\\
=&0\,.
\end{split}\end{align}
In total, the equation of motion means 
\begin{eqnarray}
\Ee = \Eo = 0\,. \label{eo1}
\end{eqnarray}

\subsection{Evaluating the flatness condition for (\ref{eq:Lax-case2})}

Next, let us examine the flatness condition for the Lax pair (\ref{eq:Lax-case2}).  

\medskip

It can be rewritten in terms of $\Se$ and $\So$ as
\begin{align}
\cL^{\sfR}_{\pm}
=&\frac{1}{1\pm \zL}
\left[\left(1\mp\frac{\eta\zL(\eta\pm R)}{1\mp\eta R} \right)\Se j_{\pm}
+ \sqrt{1+\eta^2 \zL^2}\;\So j_{\pm}\right]\,,
\label{Lax-decompose}
\end{align}
where
\begin{align}
\zL=\frac{\tanh\zR}{\tanh\alpha}\,, \qquad -i\eta=\tanh\alpha\,.
\end{align}
It is straightforward to show the following expressions: 
\begin{align}\begin{split}
\pa_{+}\cL^{\sfR}_{-} - \pa_{-}\cL^{\sfR}_{+}=&
\frac{1}{1-\zL^2}\Bigg[
\Bigg(1+\zL\left(1+\frac{\eta(\eta-R)}{1+\eta R}\right) +\zL^2\frac{\eta(\eta-R)}{1+\eta R} \Bigg)\pa_{+}\Se j_{-} \\
&\hspace{43pt}-\Bigg(1-\zL\left(1+\frac{\eta(\eta+R)}{1-\eta R}\right) +\zL^2\frac{\eta(\eta+R)}{1-\eta R} \Bigg)\pa_{-}\Se j_{+} 
\Bigg]\\
&+ \frac{\sqrt{1+\eta^2\zL^2}}{1-\zL^2}\left\{
(1+\zL)\pa_{+} \So j_{-}-(1-\zL)\pa_{-}\So j_{+} \right\}\,,
\end{split}\label{flat-diff}\end{align}
\begin{align}\begin{split}
&\hspace{-15pt}\left[\cL^{\sfR}_{+},\cL^{\sfR}_{-}\right]\\
=&
\frac{1}{1-\zL^2}\bigg(\left[\left(1-\frac{\eta\zL(\eta+R)}{1-\eta R}\right)\Se j_{+},\sqrt{1+\eta^2\zL^2}\,\So j_{-}\right]\\
&\hspace{43pt}+\left[\sqrt{1+\eta^2\zL^2}\,\So j_{+},\left(1+\frac{\eta\zL(\eta-R)}{1+\eta R}\right)\Se j_{-}\right]\\\
&\hspace{43pt}+\left[\left(1-\frac{\eta\zL(\eta+R)}{1-\eta R}\right)\Se j_{+},\left(1+\frac{\eta\zL(\eta-R)}{1+\eta R}\right)\Se j_{-}\right]\\
&\hspace{43pt}+\left[\sqrt{1+\eta^2\zL^2}\,\So j_{+},\sqrt{1+\eta^2\zL^2}\,\So j_{-}\right]\bigg)\,.\label{flat-comm}
\end{split}\end{align}
Thus the flatness condition for the Lax pair (\ref{Lax-decompose}) is evaluated as 
\begin{align}
0=\pa_{+}\cL^{\sfR}_{-}-\pa_{-}\cL^{\sfR}_{+} 
+ \left[ \cL^{\sfR}_{+} , \cL^{\sfR}_{-} \right] 
= \frac{1}{1-\zL^2}
\left(
\cF^{(\mathrm{e})}+\sqrt{1+\eta^2\zL^2}\,\cF^{(\mathrm{o})}\right)\,,
\label{flatness-even-odd}
\end{align}
where
\begin{align}\begin{split}
\cF^{(\mathrm{e})}:=&
\Bigg(1+\zL\left(1+\frac{\eta(\eta-R)}{1+\eta R}\right) +\zL^2\frac{\eta(\eta-R)}{1+\eta R} \Bigg)\pa_{+}\Se j_{-} \\
&-\Bigg(1-\zL\left(1+\frac{\eta(\eta+R)}{1-\eta R}\right) +\zL^2\frac{\eta(\eta+R)}{1-\eta R} \Bigg)\pa_{-}\Se j_{+} \\
&+\left[\left(1-\frac{\eta\zL(\eta+R)}{1-\eta R}\right)\Se j_{+},\left(1+\frac{\eta\zL(\eta-R)}{1+\eta R}\right)\Se j_{-}\right]\\
&+\left[\sqrt{1+\eta^2\zL^2}\,\So j_{+},\sqrt{1+\eta^2\zL^2}\,\So j_{-}\right]\,,
\end{split}\label{flat-even}\\
\begin{split}
\cF^{(\mathrm{o})}:=&
(1+\zL)\pa_{+} \So j_{-}-(1-\zL)\pa_{-}\So j_{+}\\
&+\left[\left(1-\frac{\eta\zL(\eta+R)}{1-\eta R}\right)\Se j_{+},\So j_{-}\right]
+\left[\So j_{+},\left(1+\frac{\eta\zL(\eta-R)}{1+\eta R}\right)\Se j_{-}\right]\,.
\end{split}\label{flat-odd}\end{align}
In summary, the flatness condition means that 
\begin{eqnarray}
\cF^{(\mathrm{e})} = \cF^{(\mathrm{o})} = 0\,. \label{eo2}
\end{eqnarray}

\subsection{Comparison between (\ref{eo1}) and (\ref{eo2})}

The remaining task is to show the equivalence between the equation of motion 
(\ref{eo1}) and the flatness condition (\ref{eo2}) 
for arbitrary values of the spectral parameter $\zL$\,.
Note that it is sufficient to establish the relation for each of 
the odd and even parts, separately.

\medskip

To this end, we will make use of the off-shell flatness condition for each part:
\begin{align}
\Ze:=\pa_+\Se j_- - \pa_-\Se j_+ +[\Se j_+,\Se j_-] + [\So j_+, \So j_-]&=0\,,\label{off-even}\\
\Zo:=\pa_+\So j_- - \pa_-\So j_+ +[\Se j_+,\So j_-] + [\So j_+, \Se j_-]&=0\,.\label{off-odd}
\end{align}

\subsubsection*{The odd part}

The $R$-matrix of Drinfeld-Jimbo type leads to the relation
\begin{align}
\frac{1}{1-\eta^2 R^2}\So j_{\pm}=\frac{1}{1+\eta^2}\So j_{\pm}\,, \qquad R(1+R^2)j_{\pm} =0\,.
\end{align}
With these relations, we can obtain
\begin{align}\begin{split}
\Fo =&\zL\,\Eo - \zL\left(
\left[\frac{\eta^3 R(1+R^2)}{1-\eta^2 R^2} \Se j_{+},\So j_{-} \right]
+\left[\So j_{+} , \frac{\eta^3 R(1+R^2)}{1-\eta^2 R^2} \Se j_{-} \right]\right)  
\\
&+\Zo
\end{split}\no\\
=&\zL\,\Eo+\Zo=\zL\,\Eo\,. 
\end{align}
Thus it follows that
\begin{align}
\Eo=0 \quad \Longleftrightarrow \quad \Fo=0 \qquad ( ^\forall \zL\in\mathbb{C})\,.
\end{align}

\subsubsection*{The even part}

Similarly, the even part of the flatness condition can be rewritten as
\begin{align}
\Fe=&
\zL\Bigg(\Ee
-\pa_{+}\left(\frac{\eta^3 R(1+R^2)}{1-\eta^2 R^2}\Se j_{-}\right)
+\pa_{-}\left(\frac{\eta^3 R(1+R^2)}{1-\eta^2 R^2}\Se j_{+}\right) \notag \\
&\hspace{17pt}+\left[\Se j_{+}, \Se j_{-}\right]-\left[\frac{\eta^3R(1+R^2)}{1-\eta^2 R^2}\Se j_{+}, \Se j_{-}\right] \notag \\
&\hspace{17pt}-\left[\Se j_{+}, \Se j_{-}\right]-\left[\Se j_{+}, \frac{\eta^3R(1+R^2)}{1-\eta^2 R^2}\Se j_{-}\right]
\Bigg) 
\notag \\
&-\zL^2\Bigg(\eta R\left(\Ee\right)
+\eta R\left(
-\pa_{+}\left(\frac{\eta^3 R(1+R^2)}{1-\eta^2 R^2}\Se j_{-}\right)
+\pa_{-}\left(\frac{\eta^3 R(1+R^2)}{1-\eta^2 R^2}\Se j_{+}\right)
\right) \notag \\
&\hspace{28pt} -\eta^2(1+\eta^2)\operatorname{mCYBE}\left(\frac{1}{1-\eta R}\Se j_{+},\frac{1}{1+\eta R}\Se j_{-}\right)
\Bigg) \notag \\
&+(1+\zL^2\eta^2)\Ze\,,
\label{flat-even-CYBE}
\end{align}
where the new symbol, mCYBE$(X,Y)$ is defined as 
\begin{align}
\operatorname{mCYBE}\left(X,Y\right):=[R(X),R(Y)]-R([R(X),Y]+[X,R(Y)])-[X,Y]=0\,,\quad (X,Y\in\mathfrak{g})\,.
\end{align}
In deriving the term including the mCYBE in (\ref{flat-even-CYBE}), 
we have used the relation
\begin{align}
\left(\pm\frac{1+\eta^2}{1-\eta^2 R^2}+\eta R\right)j_{\pm} = \left( \pm \frac{1+\eta^2}{1-\eta R}\right)j_{\pm}
\end{align}
since the $R$-operator acts as $R\rightarrow 0,\mp i$\,.

\medskip

After all, we obtain
\begin{align}
\Fe=\zL\Ee-\zL^2\;\eta R&\left(\Ee\right) +(1+\zL^2\eta^2)\Ze
=\zL\Ee-\zL^2\;\eta R\left(\Ee\right)\,,\quad ( ^\forall \zL\in\mathbb{C})\,. 
\end{align}
This means that 
\begin{align}
\Ee=0 \quad \Longleftrightarrow \quad \Fe=0 \qquad ( ^\forall \zL\in\mathbb{C})\,.
\end{align}

\section{A speciality of $SU(2)$ }

Let us consider the $SU(2)$ case for the general discussion in Subsection \ref{relating j}.
Although there are two solutions i) $\tilde{j}=j$ and ii) $\tilde{j}=\exp(\pi\Sigma)j$\,, 
they are locally equivalent because the resulting actions in (\ref{eq:YB-sigma action2}) 
and (\ref{eq:2DactionR-double}) are equivalent up to total derivative due to a speciality of $SU(2)$\,.
Hence one may anticipate a connection between the solution i) and ii)\,. 
Indeed, this is the case. We show that the two solutions are related via 
a singular formal gauge transformation.

\medskip

The Lax pair $\cL_{\pm}^{\sfR\mathrm{i)}}(\zR)$ in (\ref{eq:rescaled-LaxS3}) may be related to 
the one $\cL_{\pm}^{\sfR\mathrm{ii)}}(\zR)$ in (\ref{FR Lax}) via a gauge transformation, 
\begin{align}
\cL_{\pm}^{\sfR\mathrm{ii)}}(\zR)=g_{(+)}^{-1}\cdot \cL_{\pm}^{\sfR\mathrm{i)}}(\zR)\cdot g_{(+)}+g_{(+)}^{-1}\partial_{\pm}g_{(+)}\label{gauge-spec}
\end{align}
where 
\begin{equation}
g_{(\pm)}= \exp(\mp i (\log \wR)\,T_3)\in \mathfrak{su}(2)^{\mathbb{C}}\,. 
\end{equation}

\medskip

Since the transformation (\ref{gauge-spec}) can be regarded as a formal gauge transformation 
from $\cL^{\sfR\mathrm{ii)}}(\zR)$ to $\cL^{\sfR\mathrm{i)}}(\zR)$ by $g^{-1}_{(\pm)}$\,, 
the boundary conditions are different for each Lax pair while the bulk equation of motion 
is preserved. To see this twist explicitly, let us express the Lax pair as
\begin{align}
\cL^{\sfR}=\bar{g}^{-1}d\bar{g}\,.
\end{align}
This expression is always possible if we allow singular gauge transformations 
because the bulk equation of motion implies that $\cL^{\sfR}$ is pure gauge.
The formal gauge transformation by $g^{-1}_{(+)}=\exp(i (\log \wR)\,T_3) $
is realized by the transformation 
\begin{equation}
\bar{g}\mapsto \bar{g}'_{(+)}=\bar{g} g_{(+)}=\bar{g}\exp(- i (\log \wR)\,T_3) 
\end{equation} 
and an extra gauge transformation (\ref{extra gauge})\,.
Recalling that the boundary conditions in (\ref{A=0 trig}) lead to 
\begin{align}
\hat{g}^{-1}\partial_{\pm}\hat{g}|_{\wR=+1}=\cL^{\sfR}_{\pm}|_{\wR=+1}=\bar{g}^{-1}\pa_{\pm}\bar{g}|_{\wR=+1}\,,\\
\hat{g}^{-1}\partial_{\pm}\hat{g}|_{\wR=-1}=\cL^{\sfR}_{\pm}|_{\wR=-1}=\bar{g}^{-1}\pa_{\pm}\bar{g}|_{\wR=-1}\,,
\end{align}
if $\bar{g}^{-1}d\bar{g}$ is a solution to the boundary equation of motion satisfying 
\begin{equation}
\bar{g}^{-1}\pa_{\pm}\bar{g}|_{\wR=+1}=\bar{g}^{-1}\pa_{\pm}\bar{g}|_{\wR=-1}\,, \qquad(\Leftrightarrow \tilde{j}=j)\,,
\end{equation} 
then $\bar{g}_{(+)}^{\prime -1}d\bar{g}_{(+)}'$ is a solution satisfying 
\begin{equation}
\bar{g}'^{-1}_{(+)}\pa_{\pm}\bar{g}'_{(+)}|_{\wR=+1}=\exp \bigl(\pi R \bigr)\, 
\bar{g}'^{-1}_{(+)}\pa_{\pm}\bar{g}'_{(+)}|_{\wR=-1}\,, \qquad(\Leftrightarrow \tilde{j}=\exp(\pi R)j)\,.
\end{equation}

\medskip

In summary, the solutions i) and ii) in subsection \ref{relating j} are not topologically equivalent, though these configurations are related by a formal gauge transformation which is singular at $\wR=0,\infty$ for the $\mathfrak{su}(2)$ case.

\section{$\eta$-deformed $SL(2,\mathbb{R})$ PCM and scaling limit}

Here, let us consider an $\eta$-deformation of $SL(2,\mathbb{R})$-PCM. 
In this case, the target space becomes a warped AdS$_3$ geometry.  
Then one may consider a scaling limit of this geometry  \cite{Kawaguchi:2012ug} 
and the 3D Schr$\ddot{\rm o}$dinger spacetime \cite{Domenico}. 
We shall revisit this scaling limit at the level of a meromorphic 1-form $\omega$\,.

\subsection{Notation and classical $r$-matrices with $\mathfrak{sl}(2,\mathbb{R})$}

We first introduce the notation of the Lie algebra $\mathfrak{sl}(2,\mathbb{R})$\,. 

\medskip 
 
Let $T_{a}(a=0,1,2)$ be the generators of $\mathfrak{sl}(2,\mathbb{R})$ 
satisfying the commutation relations
\begin{align}
    [T_a,T_b]=\varepsilon_{ab}{}^{c}T_c\,. 
\end{align}
Here $\varepsilon_{ab}{}^{c}:= \varepsilon_{abd}\,\eta^{dc}$\,, $\eta_{ab}=\text{diag}(-1,1,1)$ and the antisymmetric tensor $ \varepsilon_{abc}$ is normalized as $\varepsilon_{012}=1$\,. 

\medskip

By using the Pauli matrices $\sigma^{a}$\,, the generators can be represented by 
\begin{align}
    T_0=\frac{i}{2}\sigma^2\,,\qquad T_1=\frac{1}{2}\sigma^1\,,\qquad T_2=\frac{1}{2}\sigma^3\,. 
\end{align} 
The bracket $\langle \cdot ,\cdot  \rangle$ in the deformed action (\ref{eq:YB-sigma action2}) is replaced with the trace $\Tr$, and the generators $T^{a}$'s are normalized as
\begin{align}
    {\rm Tr}(T_aT_b)=\frac{1}{2}\eta_{ab}\,,\qquad {\rm Tr}(T_+T_-)=-1\,,
\end{align}
where we introduced the light-cone combinations $T_{\pm}=\frac{1}{\sqrt{2}}(T_0\pm T_1)$\,.

\medskip

In the $SL(2,\mathbb{R})$ case, one may consider three types of classical $r$-matrix
\begin{align}
  \text{space-like}: \quad &r_{s}=2T_2\wedge T_{0}\,,\label{def:space-r}\\
  \text{time-like}: \quad   &r_{t}=2T_2\wedge T_{1}\,,\label{def:time-r}\\
 \text{light-like}: \quad &r_{l}=2T_2\wedge T_-\,,\label{def:light-r}
\end{align}
where $r_{s}$\,, $r_{t}$\,, and $r_{t}$ are called the space-, time-, and light-like $r$-matrices, respectively.
The space(time)-like $r$-matrix $r_s (r_t)$ solves the mCYBE of (non-)split type, and the associated YB deformed AdS$_3$ is called the space(time)-like warped AdS$_{3}$ spacetime. The light-like $r$-matrix $r_l$ satisfies the hCYBE, and the associated YB deformed AdS$_{3}$ is the Schr\"odinger spacetime \cite{Matsumoto:2015jja}. The space-like and time-like cases may also be called $\eta$-deformations. 

\medskip 

As explained in \cite{Kawaguchi:2012ug}, the light-like case is realized as a scaling limit of the space(time)-like warped AdS$_{3}$ spacetime.

\subsection{A scaling limit of the $\eta$-deformed $SL(2,\mathbb{R})$ PCM}

We consider here the light-like case by taking a scaling limit of the $\eta$-deformed $SL(2,\mathbb{R})$ PCM associated with the time-like $r$-matrix $r_t$ in (\ref{def:time-r})\,.  

\medskip

Let us start with the $\eta$-deformed action,
\begin{align}
S_{\rm YB}[g]=\frac{1+\tilde{\eta}^{2}}{2}\int d\tau \wedge d\sigma\operatorname{Tr} \left(g^{-1}\partial_- g\frac{1}{1-\tilde{\eta}\,R_t}g^{-1}\partial_+ g\right) \,,
\label{eq:mYBaction}
\end{align}
where $\tilde{\eta}$ is a positive real parameter and $g\in SL(2,\mathbb{R})$\,. The $R$-operator $R_t:\mathfrak{sl}(2,\mathbb{R})\to \mathfrak{sl}(2,\mathbb{R})$ associated with the time-like 
$r$-matrix $r_t$ 
in (\ref{def:time-r}) is defined as 
\begin{align}
  R_{t}(i\,T_{1}\pm T_{2}) :=\pm i\,(i\,T_{1}\pm T_{2})\,,\qquad R_t(T_0)=0\,. 
\label{eq:Rt-action}
\end{align}
It is easy to check that $R_{t}$ satisfies the mCYBE of non-split type, 
\begin{align}
  \text{CYBE}(x,y)= \frac{1}{4}[x,y]\,,\qquad x\,, y\in \mathfrak{sl}(2,\mathbb{R})\,.
\end{align}
The above deformed action can be reproduced from the 4D CS action (\ref{4dcs}) 
with an appropriate boundary condition \cite{DLMV}.

\medskip

Then, let us consider the YB deformed action with the light-like $r$-matrix (\ref{def:light-r})  
as a scaling limit of the $\eta$-deformed action (\ref{eq:mYBaction}).
To begin with, $T_{\pm}$ appearing in the time-like $r$-matrix (\ref{def:time-r}) 
are rescaled as
\begin{align}
    T_{-}\to -\frac{\sqrt{2}\eta}{\widetilde{\eta}}T_-\,,\qquad 
    T_{+}\to -\frac{\widetilde{\eta}}{\sqrt{2}\eta}T_+\,, 
\end{align}
and the time-like $r$-matrix (\ref{def:light-r}) is rewritten as 
\begin{align}
    r_t=2T_2\wedge T_1 \to \sqrt{2}T_2\wedge \left( \frac{\sqrt{2}\eta}{\widetilde{\eta}}T_--\frac{\widetilde{\eta}}{\sqrt{2}\eta}T_+\right)\,.
\end{align}
By taking the limit
\begin{align}
  \tilde{\eta}\to 0\,,\qquad \eta=\text{fixed}\,,
\end{align}
the time-like $r$-matrix $r_t$ (\ref{def:light-r}) reduces to the light-like $r$-matrix $r_l$\,(\ref{def:light-r}),
\begin{align}
    \lim_{\tilde{\eta}\to 0}\,r_t=\frac{\eta}{\tilde{\eta}}r_l+\cO(\tilde{\eta})\,,
\end{align}
and the deformed action (\ref{eq:mYBaction}) becomes
\begin{align}
S_{\rm YB}[g]=\frac{1}{2}\int d\tau \wedge d\sigma \operatorname{Tr}\left(g^{-1}\partial_- g\frac{1}{1-\eta R_l}g^{-1}\partial_+ g\right) \,.
\end{align}
This is the YB deformed action for the light-like case.

\medskip

This scaling limit can be seen at the level of a scaling limit of the meromorphic 1-form 
$\omega$ (\ref{eq:twist-eta-z}) in the trigonometric description. 
As in \cite{Kawaguchi:2012ug}, rescale the spectral parameter $z_{ \sfR}$ as
\begin{align}
\zR= \alpha\,\tilde{z}_{\sfR}\,,
\end{align}
and take a limit $\alpha\to0$\,.
Then, the limit of $\omega$ in (\ref{eq:twist-eta-z}) leads to a new $\tilde{\omega}$ as follows: 
\begin{align}
    \tilde{\omega} := \lim_{\alpha\to 0}\,\omega=\frac{(1-\tilde{z}_{\sfR}^2)}{\tilde{z}_{\sfR}^2}\,d \tilde{z}_{\sfR}\,.
\end{align}
This $\tilde{\omega}$ is a meromorphic 1-form for the homogeneous YB deformed PCM 
(or equivalently a twist function of the deformed system for the light-like case).

\end{document}